%% file: main.tex
\documentclass{article}
\PassOptionsToPackage{numbers,compress}{natbib}
\usepackage[preprint]{neurips_2026}

\usepackage[utf8]{inputenc}
\usepackage[T1]{fontenc}
\usepackage{hyperref}
\usepackage{url}
\usepackage{booktabs}
\usepackage{amsfonts}
\usepackage{nicefrac}
\usepackage{microtype}
\usepackage{xcolor}
\usepackage{graphicx}
\usepackage{amsmath} 
\usepackage{multirow} 
\usepackage{algorithm}
\usepackage{algpseudocode}
\usepackage[nolist, nohyperlinks]{acronym}
\input{acronyms.tex}

\title{KSOS-BO: Improving Sampling in Bayesian Optimization via Kernel Sum of Squares}

\author{%
  Buqing Ou
  \\
  Department of Mechanical Engineering\\
  Carnegie Mellon University\\
  Pittsburgh, PA 15213 \\
  \texttt{buqingo@andrew.cmu.edu} \\
  \And
  Frederike Dümbgen \\
  Department of Mechanical Engineering \\
  Carnegie Mellon University \\
  Pittsburgh, PA 15213 \\
  \texttt{fdumbgen@andrew.cmu.edu} \\
}

\begin{document}

\maketitle

\begin{abstract}
  Bayesian Optimization (BO) is an effective framework for globally optimizing functions whose evaluations are expensive. It is particularly effective for optimizing functions defined over continuous domains and explicitly handles stochastic noise in evaluations. As a result, it is widely applied in areas such as hyperparameter tuning, robotics policy search, and scientific experiment design, where sample efficiency is essential. Its two-step procedure consists of model fitting followed by optimization of the acquisition function, which is often treated as a generic black-box problem despite its structured nature. In this work, we introduce KSOS-BO, a kernel-based derivative-free framework for BO acquisition optimization. KSOS-BO formulates the optimization of the acquisition function as a semidefinite program with kernel-induced representations, enabling a structured global search. Across a diverse set of benchmark functions with varying landscape properties, KSOS-BO consistently outperforms derivative-free baselines using Sobol Search, Differential Evolution, or CMA-ES to optimize the acquisition function, achieving an average regret improvement of 81.16\% on 10/15 benchmarks. In particular, KSOS-BO demonstrates strong performance in highly multimodal and unimodal but ill-conditioned functions, indicating its applicability to diverse landscape structures. Despite a higher per-iteration computational cost, it converges faster in wall-clock time with an average improvement of 93.55\% on 10/15 benchmarks, as it reaches high-quality solutions with fewer evaluations. Limitations include reduced effectiveness on functions with steep drops or plate-shaped regions.
\end{abstract}

\section{Introduction}

\ac{BO}~\cite{shahriari2015taking} is a standard approach for optimizing expensive black-box functions. It builds a surrogate model from observed data and selects new evaluation points by optimizing an acquisition function. While substantial effort has been devoted to improving surrogate models and acquisition function design~\cite{snoek2015scalable,wang2017new}, the optimization of the acquisition function itself remains a critical yet often overlooked component. In practice, this inner optimization problem is often non-convex and multimodal, rendering it challenging to solve as the dimensionality grows. As a result, the overall performance of \ac{BO} can be strongly limited by the quality of acquisition optimization~\cite{gan2021acquisition}.

Existing approaches to acquisition optimization include gradient-based methods when reliable derivatives are available~\cite{wilson2018maximizing}, as well as derivative-free optimizers such as \ac{DE}~\cite{das2010differential} and \ac{CMA-ES}~\cite{hansen2016cma}. In practical \ac{GP}-based \ac{BO}, differentiable acquisition functions such as \ac{EI} are often optimized by multi-start local refinement. In this work, we focus on the complementary derivative-free setting, where the optimizer accesses the acquisition function only through function evaluations. 

Under this derivative-free setting, existing optimizers often treat acquisition optimization as a generic black-box problem and do not explicitly exploit the structure of acquisition landscapes~\cite{eggensperger2015efficient}. Their effectiveness can degrade in challenging regimes where the acquisition function is highly non-convex or multimodal. In such cases, they may require many acquisition evaluations or exhibit sensitivity to initialization, making it difficult to reliably identify high-quality candidates under a limited computational budget.

These challenges are not unique to acquisition optimization but are inherent to a broader class of non-convex global optimization problems. In such settings, local optimization methods can become trapped in suboptimal solutions, leading to degraded performance~\cite{nocedal2006numerical}. To address this issue, \ac{SOS} and moment-based relaxations have been extensively studied as principled approaches for global optimization, providing convex surrogates that enable certifiable solutions~\cite{lasserre2001global,parrilo2003semidefinite}. However, classical \ac{SOS} methods are restricted to polynomial optimization problems~\cite{lasserre2009moments}, which limits their applicability in modern machine learning~\cite{snoek2012practical} and data-driven settings where objective functions are often non-polynomial, implicitly defined, or only accessible through function evaluations. In addition, the computational cost of SOS methods can scale unfavorably with the degree of relaxation, making them difficult to apply in high-dimensional problems.

 \ac{KernelSOS}~\cite{rudi2025finding} extends \ac{SOS} beyond polynomial representations by replacing polynomial bases with kernel functions. This formulation constructs function representations from a finite set of samples and enables global optimization using only function evaluations, thereby broadening the class of admissible objective functions. Importantly, the surrogate models commonly used in \ac{BO} induce relatively smooth acquisition functions, which align well with standard kernel choices. As a result, \ac{KernelSOS} can effectively approximate and optimize these acquisition landscapes. Moreover, it replaces explicit polynomial representations with a sample-based formulation, where the computational cost depends primarily on the number of samples, which makes it naturally suited to data-driven settings.

In this paper, we propose a new optimization framework for acquisition functions, \textbf{KSOS-BO}. We adapt \ac{KernelSOS} to the acquisition optimization problem in \ac{BO}, where the objective is typically non-convex, multimodal, and accessible only through function evaluations. This perspective enables the use of structured kernel-based relaxations to improve candidate selection under a fixed acquisition-evaluation budget. In particular, we show that although \textbf{KSOS-BO} has a significantly higher per-iteration cost than competing solvers, this cost is offset by improved solution quality, almost halving the solution time to achieve the same accuracy as the derivative-free baselines considered in this study on 10/15 benchmarks. This demonstrates that enhancing the acquisition optimization step alone yields measurable improvements in overall \ac{BO} performance. In addition, the proposed method does not require extensive tuning: the same set of hyperparameters was used across all benchmarks.

\section{Related Work}
\subsection{Sum-of-Squares and Kernel Methods}

\ac{SOS} and moment-based hierarchies have been extensively studied for global optimization and polynomial analysis. They provide a principled framework for constructing convex relaxations with global guarantees~\cite{parrilo2000structured} and have been widely applied in control~\cite{tedrake2010lqr}, estimation~\cite{yang2022certifiably}, and in many practical settings, where they enable certifiable global optima~\cite{lasserre2015introduction}.

 \ac{KernelSOS} can be viewed as a generalization of polynomial \ac{SOS} methods by replacing finite-dimensional polynomials with functions in more general \ac{RKHS}~\cite{paulsen2016introduction}. Through kernel representations~\cite{scholkopf2018learning}, this formulation naturally extends to infinite-dimensional feature spaces and allows non-parametric function representations based on finite samples. Prior work has established the effectiveness of \ac{KernelSOS} in optimal control~\cite{berthier2022infinite}, value function learning~\cite{lasserre2008nonlinear}, and complex robotics control and estimation tasks~\cite{groudiev2025sampling,wei2026global}. We extend its use to an under-explored component of \ac{BO}: acquisition function optimization, where the objective is typically non-convex and multimodal, but sufficiently smooth for most common choices of surrogate functions.

\subsection{Acquisition Optimization in Bayesian Optimization}

\ac{BO} addresses black-box optimization by iteratively fitting a surrogate model and selecting query points through an acquisition function~\cite{frazier2018bayesian}. For a fixed set of observations, the acquisition function defines a deterministic objective over the input domain. This objective is typically non-convex and multimodal, exhibiting numerous local optima~\cite{jones1998efficient}. In practice, acquisition optimization is often treated as a generic black-box optimization problem~\cite{bergstra2011algorithms}. Existing approaches include gradient-based methods, which exploit local derivative information when available~\cite{conn2009introduction}, as well as derivative-free optimizers such as \ac{DE} and \ac{CMA-ES}~\cite{das2010differential,hansen2016cma}. In low-dimensional settings, acquisition functions can also be optimized via grid search over a discretized domain~\cite{brochu2009tutorial}. While these methods are broadly applicable and easy to implement, they are not specifically designed to exploit the structural properties of acquisition functions, and may therefore suffer from inefficiency or suboptimal solutions in highly multimodal landscapes.

From a structural perspective, acquisition functions are induced by surrogate models and are inherently smooth~\cite{seeger2004gaussian}. These characteristics suggest that acquisition optimization is not an arbitrary black-box problem, but rather a structured optimization problem that can benefit from methods capable of leveraging global function representations. In contrast to purely sampling-based or population-based derivative-free approaches, we formulate acquisition optimization through \ac{KernelSOS}, which constructs a structured surrogate of the acquisition function using kernel-based representations. This enables global search over the acquisition landscape using only function evaluations, allowing it to better capture the global geometry of the acquisition function and provide a more effective mechanism for identifying high-quality query points in \ac{BO}.

\section{Method}
We consider the problem of minimizing an expensive black-box function
\begin{equation}
f: \mathcal{X} \rightarrow \mathbb{R},
\end{equation}
where $\mathcal{X} \subset \mathbb{R}^d$ is a bounded domain. The objective is to identify the global minimizer
\begin{equation}
x^\star = \arg\min_{x \in \mathcal{X}} f(x).
\end{equation}

Since direct optimization of $f$ is prohibitively expensive, \ac{BO} constructs a surrogate model based on a set of noisy observations
\begin{equation}\label{eq:obs}
\mathcal{D}_t = \{(x_i, y_i)\}_{i=1}^t, \quad y_i = f(x_i) + \epsilon_i,
\end{equation}
where $\epsilon_i \sim \mathcal{N}(0, \sigma_n^2)$ denotes observation noises.
and selects new query points by optimizing an acquisition function
\begin{equation}\label{aquisition_function}
\alpha(x \mid \mathcal{D}_t).
\end{equation}

At each iteration, \ac{BO} selects the next query point by solving an acquisition optimization problem
\begin{equation}\label{argmax_aquisition_function}
x_{t+1}
=
\arg\min_{x \in \mathcal{X}}
\alpha(x \mid \mathcal{D}_t),
\end{equation}
where $\alpha(x \mid \mathcal{D}_t)$ denotes the acquisition function induced by the current surrogate model. For acquisition functions that are conventionally maximized, such as \ac{EI}, we equivalently optimize their negated forms. For example, EI maximization is written as minimizing
\begin{equation}
\alpha_{\mathrm{EI}}(x)
=
-
\mathrm{EI}(x).
\end{equation}

For fixed $\mathcal{D}_t$, the acquisition function $\alpha(x)$ is a deterministic function over $\mathcal{X}$. The corresponding optimization problem
\begin{equation}
\min_{x \in \mathcal{X}} \alpha(x)
\end{equation}
is generally non-convex and may exhibit multiple local minima and irregular structure induced by the surrogate model. This inner problem can therefore be viewed as its own global optimization problem over a bounded domain. The performance of \ac{BO} depends critically on the ability to obtain high-quality approximate solutions to this problem at each iteration within limited time.

\subsection{SOS Formulation}

We recall the standard \ac{SOS} formulation of acquisition function $\alpha$ using Eq.~\ref{aquisition_function}. Assuming $\alpha$ admits a global minimizer over $\mathcal{X}$, the problem
\begin{equation}
c^\star = \min_{x \in \mathcal{X}} \alpha(x)
\end{equation}
can be equivalently written as
\begin{equation}
c^\star = \max_{c \in \mathbb{R}} \quad \text{s.t.} \quad \alpha(x) - c \geq 0, \quad \forall x \in \mathcal{X}.
\end{equation}
This formulation seeks the largest lower bound $c$ of $\alpha$. The constraint is infinite-dimensional, and verifying the non-negativity of $\alpha(x) - c$ is, in general, intractable. A detailed mathematical formulation is provided in Appendix~\ref{app:sos_formulation}.

\subsection{KSOS Formulation}

\begin{figure}
  \centering
  \includegraphics[width=\linewidth]{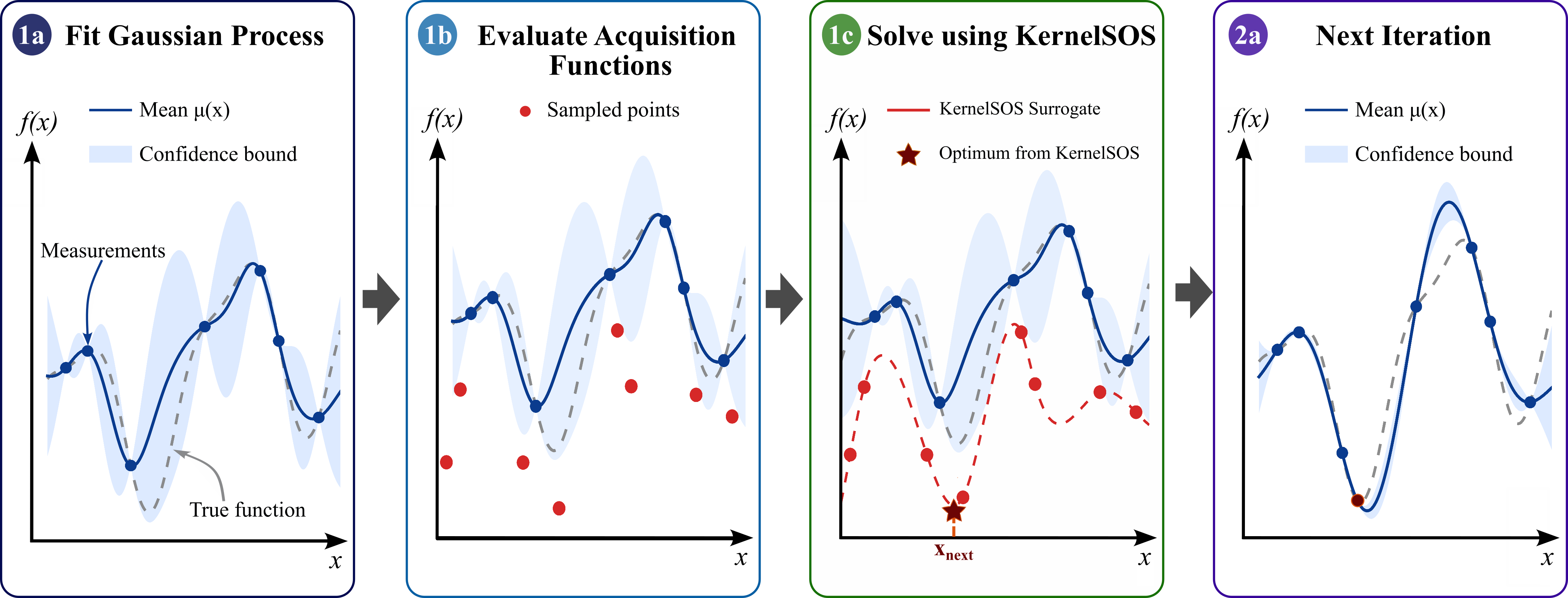}
  \caption{Overview of KSOS-BO. A Gaussian Process surrogate is fitted to the current measurement (1a), the acquisition function is evaluated at chosen points (1b), KernelSOS constructs a surrogate and minimizes it to find the next best sample (1c), and the selected point is evaluated to update the model (2a).}
  \label{fig:flow}
\end{figure}

The \ac{KernelSOS} formulation replaces the finite-dimensional polynomial basis $v_d(x)$ with functions in an \ac{RKHS} associated with a positive definite kernel $k(\cdot,\cdot)$. Rather than explicitly parameterizing the basis functions, the surrogate is represented through kernel evaluations over a finite set of sampled points $\{x_i\}_{i=1}^N$. Let $K \in \mathbb{R}^{N \times N}$ denote the kernel matrix with entries $K_{ij} = k(x_i,x_j)$.
Using the representer theorem from~\cite{marteau2020non}, the resulting finite-dimensional optimization problem can be written as
\begin{equation}\label{eq:KSOS}
\begin{aligned}
c^\star_{\mathrm{KSOS}}
=
\max_{c \in \mathbb{R},\, B \succeq 0}
\quad & c - \lambda \operatorname{Tr}(B) \\
\mathrm{s.t.}
\quad & \alpha(x_i) - c
=
\phi_i^\top B \phi_i,
\quad i = 1,\dots,N ,
\end{aligned}
\end{equation}
where
$\phi_i
=
[k(x_i,x_1),\ldots,k(x_i,x_N)]^\top ,
$
$\operatorname{Tr}(\cdot)$ denotes the matrix trace, and $\lambda \geq 0$ is a regularization parameter controlling the complexity of the surrogate. The optimization variables in Eq.~\ref{eq:KSOS} define a kernel-based surrogate representation of the shifted acquisition function rather than an explicit optimizer in the original input space. Following the recovery strategy used in the original \ac{KernelSOS} formulation, we extract a continuous candidate point as a weighted combination of the sampled points:
\begin{equation}\label{eq:ksos_minimizer}
x^\star
=
\sum_{i=1}^{N} \mu_i^\star x_i,
\end{equation}
where $\mu_i^\star$ denotes the recovery weight associated with the sampled point $x_i$. Since this recovery step is performed from the fitted KernelSOS representation rather than by directly solving a constrained optimization problem over $\mathcal{X}$, the recovered point should be interpreted as a heuristic candidate for acquisition optimization. Whenever necessary, we project $x^\star$ back to the feasible domain $\mathcal{X}$ before evaluating the expensive objective.

The quality of the recovered candidate depends on the choice of kernel, the number of sampled points $N$, and the regularization parameter $\lambda$. The kernel determines the smoothness and expressive power of the induced surrogate, where smoother kernels tend to capture broader global structure while less smooth kernels may preserve more localized variations. The number of samples $N$ controls the coverage of the acquisition landscape, and $\lambda$ balances the lower-bound objective against the complexity of the KernelSOS representation through the trace penalty on $B$.

\subsection{KernelSOS for Acquisition Optimization}

Based on the formulation above, we apply \ac{KernelSOS} to the acquisition optimization problem within each BO iteration. Given the current acquisition function $\alpha_t(x)$, we first construct a finite set of candidate points $\{x_i\}_{i=1}^N \subset \mathcal{X}$ and evaluate $\alpha_t(x_i)$ at these points. We then solve Eq.~\ref{eq:KSOS} using these acquisition values and recover a candidate optimizer using Eq.~\ref{eq:ksos_minimizer}. The recovered point is used as the next BO query point. Compared with purely sampling-based acquisition optimizers, this procedure does not simply select the best point among the sampled candidates. Instead, it fits a structured KernelSOS representation from the sampled acquisition values and uses this representation to recover a continuous candidate. This allows KSOS-BO to exploit the global structure of the acquisition landscape under a fixed acquisition-evaluation budget.
\subsection{Discussion}

The proposed method differs from standard acquisition optimizers in two key aspects. First, it constructs a global surrogate using kernel-based structure, rather than relying on local updates or samples only. Second, it performs global optimization of this surrogate via a semidefinite program, providing a principled alternative to heuristic search. Putting everything together, Figure~\ref{fig:flow} and Algorithm~\ref{alg:ksos_bo} outline the KSOS-BO algorithm. The technical implementation details of KSOS-BO are stated in Appendix~\ref{app:env_setup}.

\begin{algorithm}[t]
\caption{KSOS-BO (KernelSOS-Based Acquisition Optimization for Bayesian Optimization)}
\label{alg:ksos_bo}
\begin{algorithmic}[1]
\Require Initial dataset $\mathcal{D}_0$, acquisition $\alpha(\cdot)$, iterations $T$, sample budget $N$
\Ensure Best observed solution

\For{$t = 1,\dots,T$} \Comment{Bayesian Optimization iteration}
    \State Fit surrogate $\mathcal{M}_t$ on $\mathcal{D}_{t-1}$ and define $\alpha_t(x)$
    \State Sample $\{x_i\}_{i=1}^N$, evaluate $\{\alpha_t(x_i)\}_{i=1}^N$, and form kernel matrix $K$
    \State Solve \ac{KernelSOS} problem Eq.~\ref{eq:KSOS} and form minimizer Eq.~\ref{eq:ksos_minimizer}
    \State Evaluate $y_t=f(x_t)$ and update $\mathcal{D}_t=\mathcal{D}_{t-1}\cup\{(x_t,y_t)\}$
\EndFor

\State \textbf{return} best solution in $\mathcal{D}_T$
\end{algorithmic}
\end{algorithm}

\section{Experimental Setup}

\subsection{Acquisition Optimization Setup}

We adopt a standard \ac{BO} framework with a \ac{GP} surrogate model. At each iteration, the next query point is selected by optimizing an acquisition function. In our experiments, we use \ac{EI} as the acquisition function; its mathematical formulation is provided in Appendix~\ref{app:acquisition_functions}. All methods share the same surrogate model and acquisition function; the only difference lies in how the acquisition function is optimized.

We compare the proposed KSOS-BO optimizer with three representative derivative-free acquisition optimizers under a fixed acquisition-evaluation budget. Sobol Search performs Sobol sampling on the domain and selects the candidate with the highest value of the acquisition function, serving as a simple baseline under a fixed evaluation budget. \ac{CMA-ES} and \ac{DE} are population-based optimizers that iteratively refine candidate solutions by stochastic search. Details on the hyperparameter choices for these population-based optimizers are provided in Appendix~\ref{app:para_choice_es}, the ablation study of population size in \ac{CMA-ES} is provided in Appendix~\ref{app:cmaes} and the ablation study of population size and population diversity in \ac{DE} is provided in Appendix~\ref{app:de}. All methods use the same evaluation budget of 128 acquisition evaluations per iteration.

\subsection{Evaluation Protocol}
Each \ac{BO} run with different acquisition optimizers is repeated over 5 random seeds. We report the mean performance together with 95\% confidence intervals. At each iteration, the performance of different optimizers is measured by simple regret, defined as the gap between the true optimal value and the best value over the iterations. We assess KSOS-BO on a diverse set of benchmark functions with varying structural properties, using all suitable benchmarks from~\cite{simulationlib}.

\subsection{Implementation Details}

We follow a standard \ac{BO} setup with a \ac{GP} surrogate using a Mat\'ern kernel ($\nu=2.5$) with additive white noise. The optimization is initialized with $N_{\mathrm{init}}=12$ points and run for $400$ iterations, averaged over $5$ random seeds. We use \ac{EI} as the acquisition function with $\xi=0.01$. To ensure a fair comparison, all acquisition optimizers are given the same evaluation budget of $128$ function evaluations per iteration. Sobol sampling is used to generate candidate points for both Sobol Search and KSOS-BO. For KSOS-BO, we use a Gaussian kernel and set the search radius to $0.5$ times the domain range. The smoothing parameter is scaled as $\sigma = \lambda \cdot r / n^{1/d}$ with $\lambda=2$. Detailed configurations are provided in Appendix~\ref{app:env_setup}.

\section{Results}
\begin{table}[t]
\caption{Summary of KSOS-BO performance across benchmark functions. The table reports the relative ranking and percentage improvement compared to the second-best optimizer.}
\label{tab:ksos_summary}
\centering
\small 
\begin{tabular}{llcccc}
\toprule
\textbf{Type} & \textbf{Function} & \textbf{KSOS-BO Rank} & \textbf{Improvement (\%)} \\
\midrule

\multirow{5}{*}{Multimodal}
 & Ackley        & \textbf{1} & 99.46\\
 & Rastrigin     & \textbf{1} & 71.26\\
 & Levy        & \textbf{1}    & 39.47\\
 & Griewank        & \textbf{1}    & 99.58\\
 & Schwefel        & 4    & -22.61\\

\midrule

\multirow{1}{*}{Steep Drops}
 & Michalewicz    & 4   & -15.03\\

\midrule

\multirow{4}{*}{Bowl Shaped}
 & Rotated Hyper Ellipsoid        & \textbf{1}   & 99.24\\
 & Sphere     & 2    & -188.50\\
 & Sum of Different Powers      & \textbf{1}   & 99.74\\
 & Trid      & \textbf{1}  & 94.43\\

\midrule

\multirow{1}{*}{Plate Shaped}
 & Zakharov    & 4  & -360.00\\

\midrule

\multirow{2}{*}{Valley Shaped}
 & Dixon Price    & \textbf{1}   & 99.97\\
 & Rosenbrock    & \textbf{1}  & 92.20\\

\midrule

\multirow{2}{*}{Other}
 & Powell    & \textbf{1}   & 86.21\\
 & Styblinski Tang    & 4  & -23.04 \\

\bottomrule
\end{tabular}
\end{table}

\begin{figure}
  \centering
  \includegraphics[width=\linewidth]{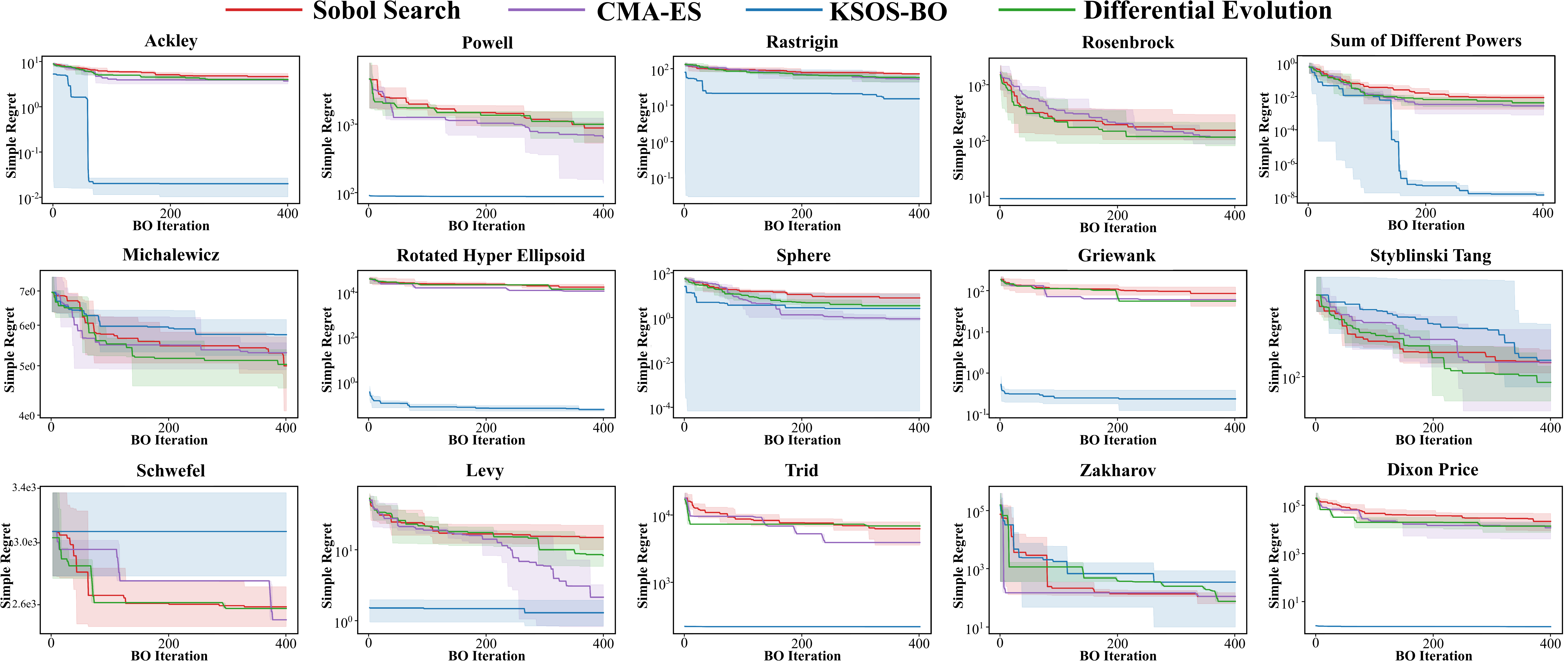}
  \caption{Convergence comparison on 15 benchmarks in 10 dimensions using \ac{EI} over 400 iterations. The curves show the simple regret over 5 random seeds on a logarithmic scale, with shaded regions indicating the 95\% confidence interval.}
  \label{fig:convergence_main}
\end{figure}

\subsection{Overall Optimization Performance}
 A quantitative summary is provided in Table~\ref{tab:ksos_summary}, and the convergence results are shown in Fig.~\ref{fig:convergence_main}. The reported improvement is computed based on the final aggregated regret. Specifically, we define
\begin{equation}
\mathrm{Improvement}(\%)
=
\frac{
r_{\mathrm{ref}} - r_{K}
}{
|r_{\mathrm{ref}}|
}
\times 100.
\end{equation}
$r_K$ denotes the final regret of KSOS-BO, and the reference value $r_{\mathrm{ref}}$ is defined as the final regret of the second-best optimizer when KSOS-BO ranks first, and as the final regret of the best optimizer otherwise. Consequently, a positive value indicates that KSOS-BO outperforms the strongest competing method, while a negative value indicates underperformance relative to the best optimizer.

In general, KSOS-BO achieves the best performance on 10 of 15 benchmarks, consistently ranking first and yielding substantial improvements over competing methods. In particular, KSOS-BO performs strongly on multimodal functions such as Ackley, Rastrigin, Levy, and Griewank, as well as on bowl-shaped functions including Sum of Different Powers, Rotated Hyper Ellipsoid, and Trid. These results indicate that KSOS-BO effectively solves both highly multimodal landscapes and smooth unimodal objectives. On smooth but curved functions such as Rosenbrock and Dixon Price, KSOS-BO attains high-quality solutions early in the optimization process with negligible variance and subsequently maintains stable performance in the identified low-regret region. In contrast,  KSOS-BO shows weaker performance on functions with steep drops or plate-shaped structures, such as Styblinski Tang, Michalewicz and Zakharov. These cases suggest that  KSOS-BO may struggle when the objective exhibits extreme curvature or overly flat regions, where the kernel-based representation becomes less adequate.

Overall, the results demonstrate that  KSOS-BO is effective in identifying high-quality solutions for acquisition optimization.

\begin{figure}
  \centering
  \includegraphics[width=\linewidth]{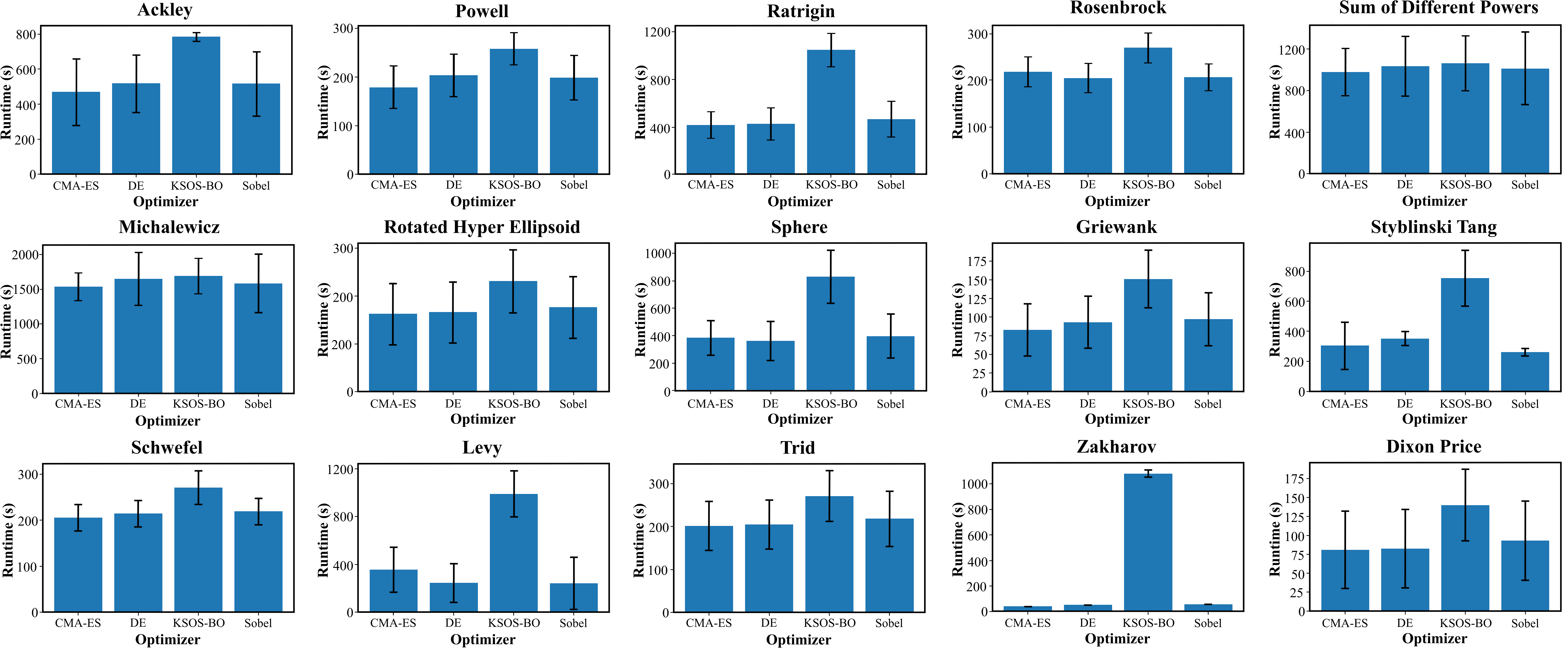}
  \caption{Average runtime over 400 iterations for four acquisition optimizers across 6 benchmark functions in 10 dimensions (Powell in 12 dimensions). Each bar shows the mean runtime over 5 random seeds, with error bars indicating one standard deviation.}
\label{fig:runtime}
\end{figure}

\subsection{Runtime and Computational Efficiency}

The proposed method consistently incurs higher computational cost compared to Sobol Search, \ac{DE}, and \ac{CMA-ES} due to the semidefinite optimization step involved in KSOS-BO. This overhead is clearly observed across all benchmarks in Fig.~\ref{fig:runtime}, where KSOS-BO requires more wall-clock time for a fixed number of iterations.

The additional cost is more pronounced on highly multimodal functions such as Ackley and Rastrigin, where the SDP takes longer to converge. In contrast, on smoother or lower-curvature functions such as Powell, Rosenbrock and Sum of Different Powers, the runtime gap is less significant. Despite this overhead, the runtime of KSOS-BO remains within the same order of magnitude as the competing methods on 14/15 tasks. This suggests that the additional cost introduced by KSOS-BO is moderate and scales predictably with the difficulty of the problem. Combined with the significant performance gains observed in Fig.~\ref{fig:convergence_main}, this trade-off between computational cost and optimization quality is favorable in practice. This is emphasized by the next section's analysis.

\subsection{Convergence Performance over Time}

\begin{figure}
  \centering
  \includegraphics[width=\linewidth]{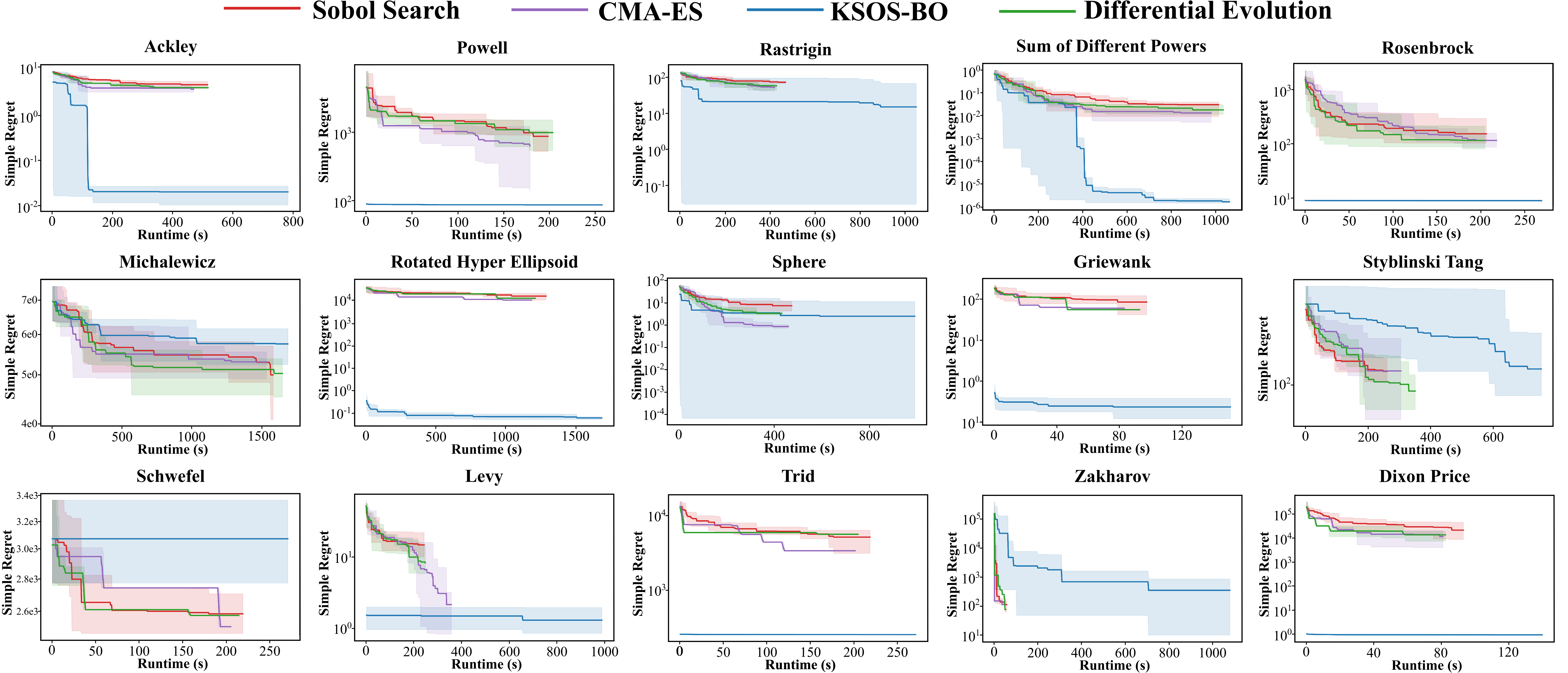}
  \caption{Convergence comparison on 15 benchmarks in 10 dimensions using \ac{EI} over wall-clock time. The curves show the simple regret over 5 random seeds on a logarithmic scale, with shaded regions indicating the 95\% confidence interval.}
\label{fig:convergence_over_time}
\end{figure}

We evaluate the efficiency of different acquisition optimizers in terms of wall-clock time. Figure~\ref{fig:convergence_over_time} shows the simple regret reduction over runtime across 15 benchmark functions, averaged over $5$ random seeds for $400$ iterations. Due to space limitations, the quantitative summary table is reported in Appendix~\ref{app:runtime_table}.

Although KSOS-BO incurs higher computational cost per iteration, it consistently achieves faster convergence in terms of wall-clock time. Across many benchmarks, KSOS-BO reaches low-regret regions significantly earlier, requiring fewer iterations to achieve comparable performance. This indicates that KSOS-BO is not only sample-efficient but also time-efficient in practice. The advantage is particularly clear on Powell, Rosenbrock, Rotated Hyper Ellipsoid, Griewank, Levy, Trid and Dixon Price, where KSOS-BO outperforms all baselines from the very beginning with negligible variance and maintains this lead throughout optimization. On multimodal functions such as Ackley and Rastrigin, KSOS-BO also exhibits faster and more stable progress, while competing methods converge more slowly and plateau at higher regret levels.

Overall, these results show that the higher per-iteration cost of KSOS-BO is offset by substantially improved convergence speed. When measured in wall-clock time, KSOS-BO consistently reaches better solutions faster on most benchmarks, making it a more effective optimizer in practice.

\subsection{\ac{KernelSOS} Surrogate Approximation of \ac{EI}}

\begin{figure}
  \centering
  \includegraphics[width=0.76\linewidth]{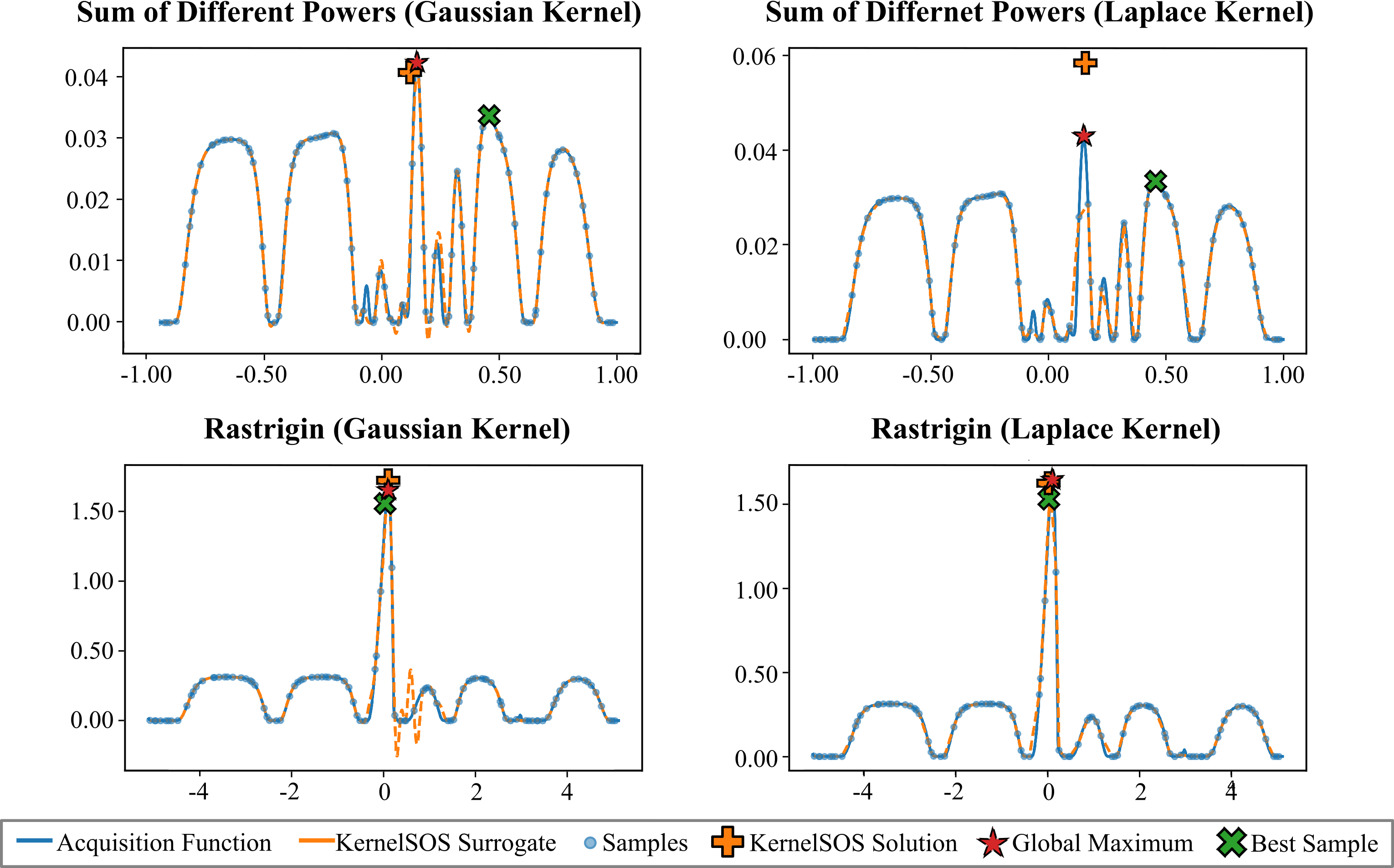}
  \caption{
Visualization of \ac{KernelSOS} surrogate approximation on 1D \ac{EI} after 9 \ac{BO} steps.
}
\label{fig:ksos_surrogate_main}
\end{figure}

To understand the behavior of KSOS-BO, we examine how its surrogate approximates the true \ac{EI} acquisition landscape in $1$D during \ac{BO}. Figure~\ref{fig:ksos_surrogate_main} shows the acquisition functions after $9$ \ac{BO} steps, together with the \ac{KernelSOS} surrogate fitted from $128$ Sobol samples using Gaussian and Laplace kernels. With both Gaussian and Laplace kernels, the surrogate closely matches the true acquisition curves on Sum of Different Powers and Rastrigin. The overall shape of the acquisition landscape, including multiple local peaks and the dominant global maximum, is accurately captured, leading to strong alignment across the domain. Even for the highly localized and peaked structure of \ac{EI}, the surrogate successfully identifies the key maxima, with only minor discrepancies in relatively flat regions.

Overall, these results show that KSOS-BO can approximate complex acquisition landscapes from a limited number of evaluations. The Gaussian kernel produces smoother and more globally consistent approximations, while the Laplace kernel provides higher local resolution and better captures sharper variations. Despite these differences, both kernels reliably identify the optimal query, indicating that KSOS-BO is robust to the choice of kernel.

\section{Conclusion}
We presented KSOS-BO, a kernel-based derivative-free method for acquisition optimization in \ac{BO}. Using semidefinite programming and kernel-induced representations, KSOS-BO performs structured global search over complex acquisition landscapes. Across diverse benchmarks, it consistently achieves strong performance, ranking first by a large margin on 10/15 benchmarks, outperforming Sobol Search, \ac{DE}, and \ac{CMA-ES}. KSOS-BO performs well on both multimodal and smooth unimodal problems, indicating its ability to capture different landscape structures. Despite higher per-iteration cost, it converges faster in wall-clock time on most tasks, reaching high-quality solutions with fewer evaluations. This highlights the impact of improving the acquisition optimization step on overall Bayesian optimization performance. However, KSOS-BO is less effective on functions with extreme curvature or flat regions, where the kernel representation becomes less expressive. The semidefinite optimization step also introduces additional computational overhead. Future work will focus on improving the efficiency and scalability of KSOS-BO, including more efficient optimization formulations and appropriate kernels. Another important direction is to compare KSOS-BO with strong multi-start gradient-based acquisition optimizers and to study hybrid strategies that use KSOS-BO for global candidate generation followed by local refinement. Evaluating the method in higher-dimensional settings and other acquisition functions is another important direction. Finally, KSOS-BO should be tested in advanced \ac{BO} pipelines, including constrained and multi-objective settings.

\newpage
\appendix

\section{THEORETICAL BACKGROUND}

\subsection{Gaussian Process}

A Gaussian process (GP) is a collection of random variables $\{f(x)\}_{x \in \mathcal{X}}$ such that any finite subset follows a joint Gaussian distribution. A GP is fully specified by a mean function $m(x)$ and a covariance function $k(x,x')$, and is denoted as~\cite{williams1995gaussian}
\begin{equation}
f(x) \sim \mathcal{GP}\big(m(x), k(x,x')\big).
\end{equation}
For any inputs $x_1, \dots, x_n$, the corresponding function values satisfy
\begin{equation}
\big(f(x_1), \dots, f(x_n)\big)^\top \sim \mathcal{N}\big(\mu, K\big),
\end{equation}
where $\mu_i = m(x_i)$ and $K_{ij} = k(x_i, x_j)$.

Given observations $y = (y_1, \dots, y_n)^\top$ at inputs $X = \{x_i\}_{i=1}^n$, with Gaussian noise $y_i = f(x_i) + \epsilon_i$ and $\epsilon_i \sim \mathcal{N}(0, \sigma^2)$, the posterior distribution at a test point $x_*$ remains Gaussian with mean and variance
\begin{equation}
\mu(x_*) = k_*^\top (K + \sigma^2 I)^{-1} y, \quad
\sigma^2(x_*) = k(x_*, x_*) - k_*^\top (K + \sigma^2 I)^{-1} k_*,
\end{equation}
where $k_* = (k(x_*, x_1), \dots, k(x_*, x_n))^\top$.

The covariance function $k(x,x')$ is required to be symmetric and positive definite, which ensures that the kernel matrix $K$ is positive semidefinite for any finite set of inputs.

\subsection{Acquisition Functions}\label{app:acquisition_functions}
We consider two commonly used acquisition functions in \ac{BO}: \ac{EI} and \ac{LCB}.

\ac{EI} is defined as
\begin{equation}
\alpha_{\text{EI}}(x) = (f_{\text{best}} - \mu(x) - \xi)\,\Phi\!\left(\frac{f_{\text{best}} - \mu(x) - \xi}{\sigma(x)}\right)
+ \sigma(x)\,\phi\!\left(\frac{f_{\text{best}} - \mu(x) - \xi}{\sigma(x)}\right),
\end{equation}
where $\mu(x)$ and $\sigma(x)$ denote the posterior mean and standard deviation of the surrogate model, $f_{\text{best}}$ is the best observed value, $\Phi(\cdot)$ and $\phi(\cdot)$ are the standard normal CDF and PDF, and $\xi \geq 0$ represents the exploration parameter.

{LCB} is defined as
\begin{equation}
\alpha_{\text{LCB}}(x) = \mu(x) - \beta \sigma(x),
\end{equation}
where $\beta > 0$ is the exploration parameter controlling the trade-off between exploration and exploitation.

\subsection{Sum-of-Squares Relaxations for Polynomial Optimization}
\label{app:sos_formulation}

We briefly review the \ac{SOS} framework for global polynomial optimization, which forms the foundation of the \ac{KernelSOS} extension used in this work~\cite{parrilo2000structured}.

\paragraph{Global optimization via nonnegativity.}
Consider the unconstrained global minimization problem
\begin{equation}
f^* = \min_{x \in \mathcal{X}} f(x).
\end{equation}
This problem can be equivalently reformulated as the search for the largest scalar lower bound
\begin{equation}
c^* = \max_{c \in \mathbb{R}} \quad \text{s.t.} \quad f(x) - c \geq 0, \quad \forall x \in \mathcal{X},
\end{equation}
where the constraint enforces global nonnegativity of the shifted function $f(x) - c$. However, verifying global nonnegativity of a multivariate polynomial is intractable in general.

\paragraph{Sum-of-squares relaxation.}
A sufficient condition for nonnegativity is that the polynomial admits a \ac{SOS} decomposition. Specifically, a polynomial $p(x)$ is said to be \ac{SOS} if there exist polynomials $\{q_i(x)\}_{i=1}^m$ such that
\begin{equation}
p(x) = \sum_{i=1}^m q_i^2(x).
\end{equation}
This condition ensures $p(x) \geq 0$ for all $x$, and leads to the \ac{SOS} relaxation of the global optimization problem:
\begin{equation}
c^*_{\mathrm{SOS}} = \max_{c \in \mathbb{R}} \quad \text{s.t.} \quad f(x) - c \quad \text{is} \quad\mathrm{SOS}.
\end{equation}

Since the \ac{SOS} condition is sufficient but not necessary for nonnegativity, the relaxation provides a lower bound
\begin{equation}
c^*_{\mathrm{SOS}} \leq c^*.
\end{equation}

\paragraph{Semidefinite representation.}
A key property of \ac{SOS} polynomials is that they admit a semidefinite representation. Let $v_d(x)$ denote the vector of monomials up to degree $d$. Then a polynomial $p(x)$ of degree $2d$ is \ac{SOS} if and only if there exists a positive semidefinite matrix $Q \succeq 0$ such that
\begin{equation}
p(x) = v_d(x)^\top Q v_d(x).
\end{equation}
This equivalence transforms the constraint \ac{SOS} into a constraint \ac{LMI}, allowing the relaxed problem to be solved by semidefinite programming.

\paragraph{Hierarchy of relaxations.}
The \ac{SOS} relaxation can be systematically tightened by increasing the degree of the monomial basis $v_d(x)$. This leads to a hierarchy of semidefinite programs whose optimal values converge to the true global optimum under mild conditions~\cite{parrilo2000structured}.

In particular, for polynomial optimization problems, this hierarchy provides a sequence of convex relaxations that approximate the original nonconvex problem with increasing accuracy and, in some cases, recover the exact global optimum at a finite level.

This polynomial formulation \ac{SOS} provides the foundation for the kernelized extension presented in the next subsection, where the finite-dimensional monomial basis is replaced by functions in a \ac{RKHS}.

\subsection{Kernel Sum-of-Squares in Reproducing Kernel Hilbert Spaces}

The polynomial \ac{SOS} formulation can be extended to an infinite-dimensional setting by considering functions in a reproducing kernel Hilbert space (\ac{RKHS})~\cite{marteau2020non}. This extension replaces explicit polynomial bases with kernel-induced feature maps, allowing nonnegative functions to be represented through positive operators in an \ac{RKHS}.

\paragraph{RKHS and function representation.}
Let $\mathcal{H}$ be an \ac{RKHS} associated with a positive definite kernel $k(\cdot,\cdot)$. By the reproducing property, any function $g \in \mathcal{H}$ satisfies
\begin{equation}
g(x)
=
\langle g, k(x,\cdot) \rangle_{\mathcal{H}} .
\end{equation}
Given a finite set of sample points $\{x_i\}_{i=1}^N$, the representer theorem~\cite{rudi2025finding} implies that solutions to regularized optimization problems over $\mathcal{H}$ can be expressed as finite kernel expansions of the form
\begin{equation}
g(x)
=
\sum_{i=1}^{N} \alpha_i k(x,x_i),
\end{equation}
for some coefficients $\alpha \in \mathbb{R}^N$.

\paragraph{Kernelized sum-of-squares representation.}
In classical \ac{SOS} optimization, a nonnegative function is represented as a sum of squared functions. KernelSOS generalizes this idea by replacing finite-dimensional polynomial bases with functions in an \ac{RKHS}. Let $\phi(x)$ denote the feature map associated with the kernel $k$, such that
\begin{equation}
k(x,x')
=
\langle \phi(x), \phi(x') \rangle_{\mathcal{H}} .
\end{equation}
A function $s(x)$ admits a KernelSOS representation if it can be written as
\begin{equation}
s(x)
=
\langle \phi(x), B \phi(x) \rangle_{\mathcal{H}},
\end{equation}
where $B \succeq 0$ is a positive semidefinite operator acting on the feature space. In the infinite-dimensional formulation, this construction ensures
\begin{equation}
s(x) \geq 0,
\quad \forall x \in \mathcal{X}.
\end{equation}
Thus, if a function $f(x)-c$ admits such a representation, then $c$ is a valid lower bound of $f$ over the domain $\mathcal{X}$.

\paragraph{Infinite-dimensional lower-bound formulation.}
Using this representation, the KernelSOS relaxation of a minimization problem can be written as
\begin{equation}
\begin{aligned}
\max_{c \in \mathbb{R},\, B \succeq 0}
\quad & c - \lambda \operatorname{Tr}(B) \\
\mathrm{s.t.}
\quad & f(x) - c
=
\langle \phi(x), B\phi(x) \rangle_{\mathcal{H}},
\quad \forall x \in \mathcal{X},
\end{aligned}
\end{equation}
where $\lambda \geq 0$ is a regularization parameter. Since $B \succeq 0$, the right-hand side is nonnegative for every $x \in \mathcal{X}$, and therefore the constraint implies
\begin{equation}
f(x) - c \geq 0,
\quad \forall x \in \mathcal{X}.
\end{equation}
The infinite-dimensional program therefore has a lower-bound interpretation: the scalar $c$ is optimized subject to being no larger than $f(x)$ over the entire domain. Under the assumptions studied in the original KernelSOS analysis, including sufficient smoothness of $f$ and compatibility with the chosen \ac{RKHS}, this formulation provides a principled relaxation for global minimization~\cite{marteau2020non}.

\paragraph{Connection to finite-dimensional approximation.}
The constraint over all $x \in \mathcal{X}$ is generally intractable. Following the original KernelSOS formulation, we approximate it using a finite set of sampled points $\{x_i\}_{i=1}^{N}$. Evaluating the feature map at these points and applying the representer theorem yields a finite-dimensional semidefinite program of the form used in the main text.

This finite-dimensional problem should be interpreted as a sampled approximation of the infinite-dimensional KernelSOS program. In particular, the practical formulation enforces the KernelSOS equality constraints only at the sampled points,
\begin{equation}
f(x_i) - c
=
\phi_i^\top B \phi_i,
\quad i = 1,\ldots,N,
\end{equation}
where $\phi_i = [k(x_i,x_1),\ldots,k(x_i,x_N)]^\top$. Therefore, the finite formulation does not by itself certify exact global nonnegativity over the entire domain. Instead, it provides a kernel-based sampled relaxation whose quality depends on the kernel, the sample distribution, the number of sampled points, and the regularization parameter.

The original KernelSOS analysis shows that, for sufficiently smooth functions and suitable kernels, the finite sampled formulation approximates the infinite-dimensional lower-bound problem, with the approximation quality controlled by the coverage of the sampled points and the smoothness of the function. In our setting, the function being optimized by KernelSOS is the acquisition objective. Common acquisition functions induced by Gaussian process surrogates, such as \ac{EI} and \ac{LCB}, are typically smooth functions when the posterior mean and variance are smooth. This makes the \ac{RKHS}-based approximation assumption reasonable in practice. Nevertheless, our finite-dimensional implementation should be understood as a structured sampled approximation for acquisition optimization, rather than as a formal certificate of the exact global optimum of the acquisition function.

\subsection{Benchmark Functions}
\label{app:benchmarks}

We evaluate all optimization methods on a diverse set of $15$ standard benchmark functions that cover a range of landscape characteristics, including multimodal functions with many local minima, Bowl-shaped, Valley-shaped, Plate-Shaped, and steep-drop landscapes. For each function, we provide its analytical form, search domain, and known global optimum.

\subsubsection{Multimodal}

\paragraph{Ackley}
\begin{equation}
f(x) = -20 \exp\left(-0.2 \sqrt{\frac{1}{d}\sum_{i=1}^{d} x_i^2}\right)
- \exp\left(\frac{1}{d}\sum_{i=1}^{d} \cos(2\pi x_i)\right)
+ 20 + e
\end{equation}
Domain: $x_i \in [-5,5]$ \\
Global minimum: $f(x^*) = 0$ at $x^* = (0,\dots,0)$

\paragraph{Rastrigin}
\begin{equation}
f(x) = 10d + \sum_{i=1}^{d} \left(x_i^2 - 10 \cos(2\pi x_i)\right)
\end{equation}
Domain: $x_i \in [-5.12,5.12]$ \\
Global minimum: $f(x^*) = 0$ at $x^* = (0,\dots,0)$

\paragraph{Levy}
\begin{equation}
w_i = 1 + \frac{x_i - 1}{4}
\end{equation}
\begin{equation}
f(x) = \sin^2(\pi w_1) + \sum_{i=1}^{d-1} (w_i - 1)^2 \left(1 + 10 \sin^2(\pi w_i + 1)\right)
+ (w_d - 1)^2 (1 + \sin^2(2\pi w_d))
\end{equation}
Domain: $x_i \in [-10,10]$ \\
Global minimum: $f(x^*) = 0$ at $x^* = (1,\dots,1)$

\paragraph{Griewank}
\begin{equation}
f(x) = \frac{1}{4000}\sum_{i=1}^{d} x_i^2 - \prod_{i=1}^{d} \cos\left(\frac{x_i}{\sqrt{i}}\right) + 1
\end{equation}
Domain: $x_i \in [-600,600]$ \\
Global minimum: $f(x^*) = 0$ at $x^* = (0,\dots,0)$

\paragraph{Schwefel}
\begin{equation}
f(x) = 418.9829d - \sum_{i=1}^{d} x_i \sin(\sqrt{|x_i|})
\end{equation}
Domain: $x_i \in [-500,500]$ \\
Global minimum: $f(x^*) = 0$ at $x_i = 420.9687$

\subsubsection{Bowl-Shaped Functions}

\paragraph{Sphere}
\begin{equation}
f(x) = \sum_{i=1}^{d} x_i^2
\end{equation}
Domain: $x_i \in [-5,5]$ \\
Global minimum: $f(x^*) = 0$ at $x^* = 0$

\paragraph{Rotated Hyper Ellipsoid}
\begin{equation}
f(x) = \sum_{i=1}^{d} \sum_{j=1}^{i} x_j^2
\end{equation}
Domain: $x_i \in [-5,5]$ \\
Global minimum: $f(x^*) = 0$

\paragraph{Sum of Different Powers}
\begin{equation}
f(x) = \sum_{i=1}^{d} |x_i|^{i+1}
\end{equation}
Domain: $x_i \in [-5,5]$ \\
Global minimum: $f(x^*) = 0$

\paragraph{Trid}
\begin{equation}
f(x) = \sum_{i=1}^{d} (x_i - 1)^2 - \sum_{i=2}^{d} x_i x_{i-1}
\end{equation}
Domain: $x_i \in [-5,5]$ \\
Global minimum: $f(x^*) = -\frac{d(d+4)(d-1)}{6}$

\subsubsection{Plate-Shaped Functions}

\paragraph{Zakharov}
\begin{equation}
f(x) = \sum_{i=1}^{d} x_i^2 + \left(\sum_{i=1}^{d} \frac{i}{2} x_i \right)^2 + \left(\sum_{i=1}^{d} \frac{i}{2} x_i \right)^4
\end{equation}
Domain: $x_i \in [-5,10]$ \\
Global minimum: $f(x^*) = 0$

\subsubsection{Valley-Shaped Functions}

\paragraph{Rosenbrock}
\begin{equation}
f(x) = \sum_{i=1}^{d-1} \left[100(x_{i+1} - x_i^2)^2 + (1 - x_i)^2\right]
\end{equation}
Domain: $x_i \in [-2,2]$ \\
Global minimum: $f(x^*) = 0$ at $x^* = (1,\dots,1)$

\paragraph{Dixon Price}
\begin{equation}
f(x) = (x_1 - 1)^2 + \sum_{i=2}^{d} i (2x_i^2 - x_{i-1})^2
\end{equation}
Domain: $x_i \in [-10,10]$ \\
Global minimum: $f(x^*) = 0$

\subsubsection{Steep Drops}

\paragraph{Michalewicz}
\begin{equation}
f(x) = -\sum_{i=1}^{d} \sin(x_i) \left[\sin\left(\frac{i x_i^2}{\pi}\right)\right]^{2m}
\end{equation}
Domain: $x_i \in [0,\pi]$ \\
Global minimum: depends on $d$ (e.g., $d=10$: $\approx -9.66$)

\subsubsection{Other Functions}

\paragraph{Powell}
\begin{equation}
f(x) = \sum_{k=1}^{d/4} \Big[(x_{4k-3} + 10x_{4k-2})^2 + 5(x_{4k-1} - x_{4k})^2
+ (x_{4k-2} - 2x_{4k-1})^4 + 10(x_{4k-3} - x_{4k})^4 \Big]
\end{equation}
Domain: $x_i \in [-4,5]$ \\
Global minimum: $f(x^*) = 0$

\paragraph{Styblinski Tang}
\begin{equation}
f(x) = \frac{1}{2} \sum_{i=1}^{d} (x_i^4 - 16x_i^2 + 5x_i)
\end{equation}
Domain: $x_i \in [-5,5]$ \\
Global minimum: $f(x^*) = -39.16599d$ at $x_i \approx -2.903534$

\section{Experimental Details}

\subsection{Experimental Setup Details}\label{app:env_setup}

Here we provide the  implementation details of KSOS-BO in Table~\ref{tab:exp_setup}. $N_{\text{init}}$ denotes the number of initial points evaluated before \ac{BO} begins, while $N_{\text{iters}}$ denotes the number of subsequent \ac{BO} iterations used for sequential acquisition and evaluation. $\sigma_{\text{n}}$ denotes the standard deviation of the observation noise defined in Eq.~\ref{eq:obs}. For acquisition functions, $\xi$ is the improvement threshold in \ac{EI}. In \ac{CMA-ES}, $\sigma_0$ denotes the initial sampling step size, and \textit{range} refers to the width of the search domain along each dimension. In KSOS-BO, $r$ denotes the sampling radius, defined based on the domain bounds (upper and lower). The smoothing parameter $\sigma$ is defined as $\sigma =\lambda \cdot r / n^{1/d}$, where $\lambda$ is a scaling parameter, $r$ denotes the sampling radius, $n$ is the number of samples, and $d$ is the dimension.

\begin{table}[t]
\centering
\caption{Experimental setup for KSOS-BO. All methods share the same evaluation budget and search domain unless otherwise specified.}
\label{tab:exp_setup}
\small
\begin{tabular}{l l}
\toprule
\textbf{Component} & \textbf{Configuration} \\
\midrule

\textbf{Objective Functions} 
& Standard benchmarks (Ackley, Rastrigin, Rosenbrock, etc.) \\

\textbf{Dimension} 
& $d = 10$ \\

\textbf{BO Iterations} 
& $N_{\text{init}} = 12$; $N_{\text{iters}} = 400$ \\

\textbf{Random Seeds} 
& $5$ \\

\midrule
\textbf{Gaussian Process} 
& Constant $\times$ Matern($\nu = 2.5$) + WhiteKernel \\

\textbf{Length-scale Bounds} 
& $[10^{-2}, 10^{2}]$ \\

\textbf{Observation Noise} 
& $\sigma_{\text{n}} = 0.05 \cdot \mathrm{std}(y)$ \\

\midrule
\textbf{Acquisition Function} 
& Expected Improvement (EI) \\

\textbf{EI Parameter} 
& $\xi = 0.01$ \\

\midrule
\textbf{Evaluation Budget} 
& $128$ evaluations per iteration \\

\midrule
\textbf{Sobol Search} 
& Sobol sampling over the full domain \\

\textbf{CMA-ES} 
& Population size $= 8$; 
$\sigma_0 = 0.3 \cdot \text{range}$ \\

\textbf{Differential Evolution} 
& Population size $= 2$; $5$ iterations; 
$\approx 120$ evaluations \\

\textbf{KSOS-BO (Ours)} 
& Sobol sampling; Gaussian kernel; Newton solver \\

\textbf{KSOS-BO Radius} 
& $r = 0.5 \cdot (\text{upper} - \text{lower})$ \\

\textbf{KSOS-BO Smoothing} 
& $\lambda = 2$; $\sigma = \lambda \cdot r / n^{1/d}$ \\

\bottomrule
\end{tabular}
\end{table}

\subsection{Parameter Settings for Evolutionary Acquisition Optimizers}\label{app:para_choice_es}

We provide a brief justification for the parameter choices used in evolutionary optimizers, namely \ac{CMA-ES} and \ac{DE}, under the fixed evaluation budget setting adopted in our experiments.

For \ac{CMA-ES}, we follow the standard parameterization proposed in the literature~\cite{hansen2016cma}. In particular, the population size is typically chosen as
\begin{equation}
\lambda = 4 + \lfloor 3 \log d \rfloor,
\end{equation}
where $d$ denotes the dimension of the problem. This rule suggests that the population size should increase logarithmically with dimension. However, in our setting, optimization is performed under a strict evaluation budget of $128$ evaluations per iteration. To ensure that the budget is fully utilized and comparable between optimizers, we fix the population size at $\lambda = 8$, which evenly divides the evaluation budget and yields a sufficient number of optimization generations. 

For step-size initialization, the literature~\cite{hansen2016cma} recommends setting the initial scale proportional to the search domain. Specifically, when the search domain is $[a,b]^d$, the initial step-size is chosen as
\begin{equation}
\sigma_0 = 0.3 (b - a),
\end{equation}
which ensures that the initial sampling distribution reasonably covers the search space while allowing subsequent adaptation during optimization. We adopt this rule directly in our implementation.

For \ac{DE}, we follow the standard parameterization proposed in the literature~\cite{das2010differential}. Specifically, parameter selection is less prescriptive. Unlike \ac{CMA-ES}, \ac{DE} does not provide a fixed rule for population size, and its performance is known to depend on the interaction between mutation factor $F$, crossover rate $Cr$, and population diversity. In standard settings, relatively large populations are often recommended. However, under the same strict evaluation budget of 128 function evaluations, such configurations are not feasible~\cite{das2010differential}.

To ensure a fair comparison between optimizers, we adopt a reduced population size and iteration count, specifically setting $\texttt{popsize}=2$ and $\texttt{maxiter}=5$, resulting in approximately 120 function evaluations per optimization step. Although this configuration deviates from classical large-population settings, it remains consistent with the characteristics of DE. In particular, the mutation factor $F$ controls the magnitude of the perturbations, and the crossover rate $Cr$ determines the number of modified dimensions, thus governing the effective exploration behavior. We use $\texttt{mutation}=(0.5,1)$ and $\texttt{recombination}=0.7$, which correspond to commonly adopted values that provide sufficient exploration.

Furthermore, previous studies have shown that DE inherently maintains a relatively high population variance compared to other evolutionary strategies, allowing effective exploration even with smaller populations. Therefore, reducing the population size under a fixed evaluation budget primarily trades off population diversity for additional iterative refinement, without completely sacrificing exploratory capability.


\section{Additional Studies}

\subsection{Ablation Study on \ac{LCB}}
In addition to \ac{EI} used in the main text, we further evaluate all optimizers using \ac{LCB}. Its mathematical formulation is provided in Appendix~\ref{app:acquisition_functions}. Figure~\ref{fig:LCB_convergence} shows the convergence curves (with 95\% confidence intervals over 5 seeds) on 6 representative benchmark functions. We set $\beta = 2$ as the exploration parameter in \ac{LCB}. In addition to the choice of acquisition function, all other experimental settings are identical to those in Figure~\ref{fig:convergence_main}.

In general, the qualitative behavior remains consistent with the \ac{EI} setting. KSOS-BO continues to exhibit faster convergence in early iterations and achieves lower final regret on Ackley, Powell, Rastrigin, and Rosenbrock. This suggests that the advantage of KSOS-BO is not specific to a particular acquisition function.

\begin{figure}
  \centering
  \includegraphics[width=\linewidth]{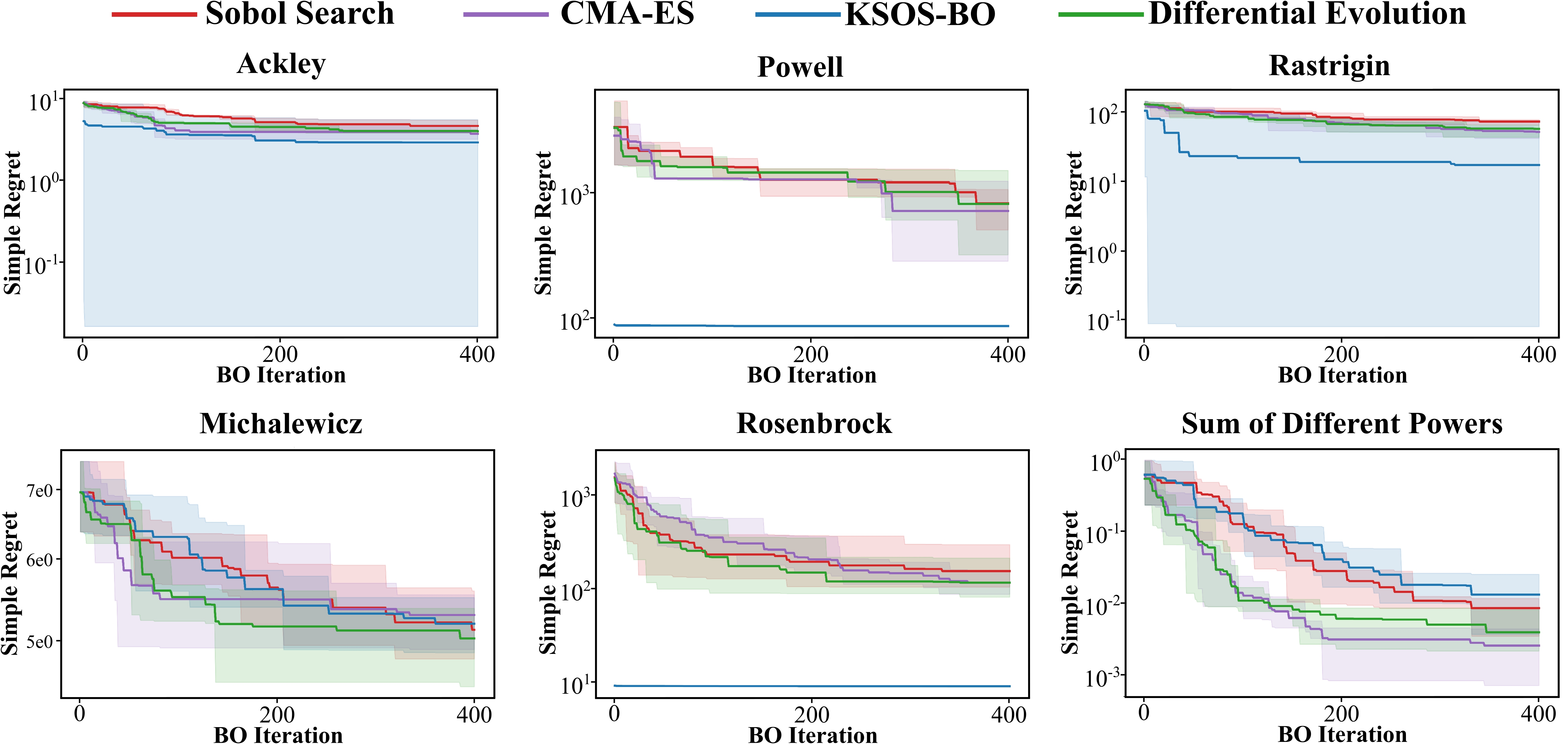}
  \caption{
Convergence comparison on 6 benchmarks in 10 dimensions using \ac{LCB} over 400 iterations. The curves show the simple regret over 5 random seeds on a logarithmic scale, with shaded regions indicating the 95\% confidence interval.
}
\label{fig:LCB_convergence}
\end{figure}

\subsection{Ablation Study on $N_{\text{init}}$}
\begin{figure}
  \centering
  \includegraphics[width=\linewidth]{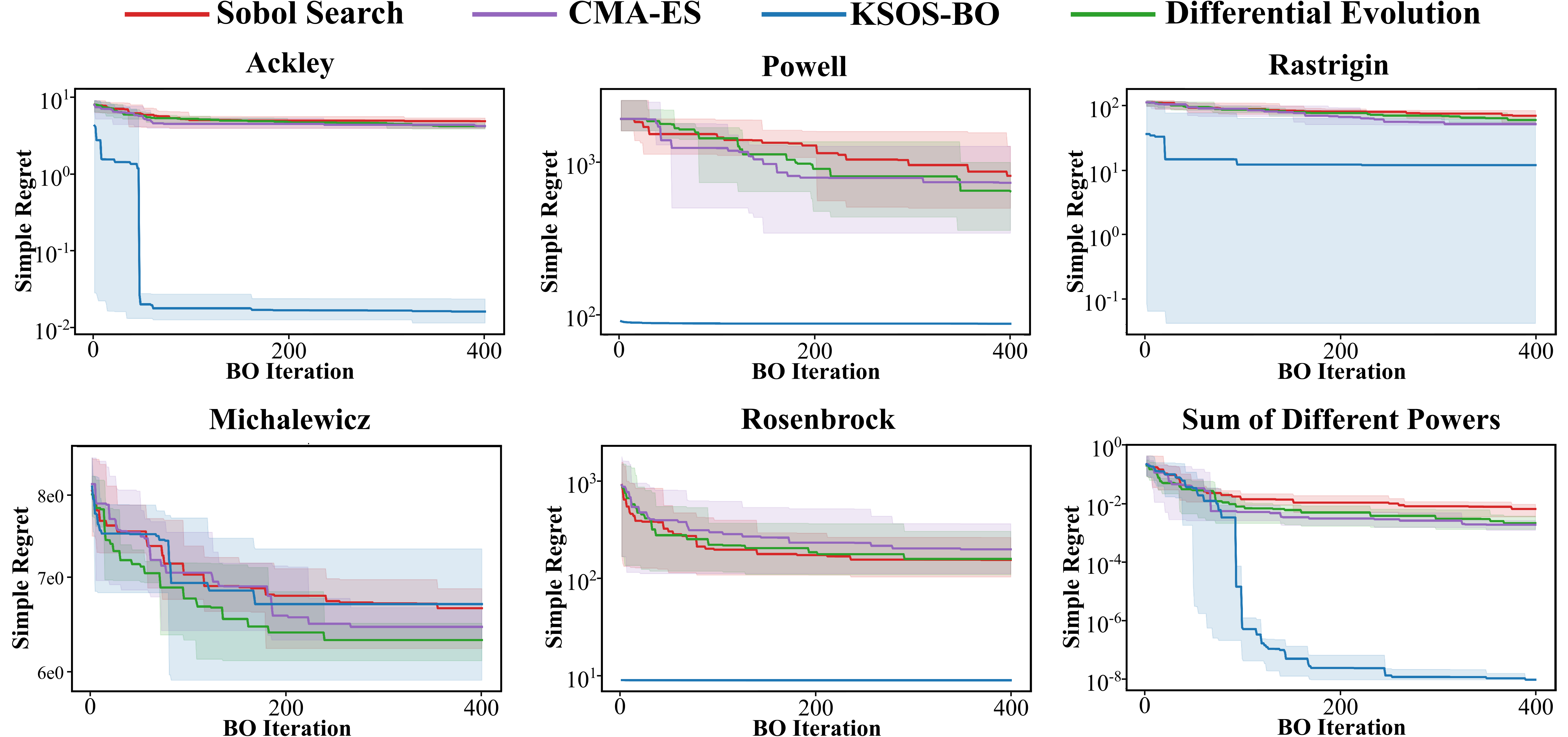}
  \caption{
   Convergence comparison on 6 10-dimensional benchmark functions with $N_{\text{init}}=32$ using \ac{EI} over 400 iterations. The curves show the simple regret across 5 random seeds on a logarithmic scale, with shaded regions indicating the 95\% confidence interval.
}
\label{fig:ninit_32_convergence}
\end{figure}

We study the influence of the initial design size $N_{\text{init}}$ on the performance of different optimizers. In \ac{BO}, $N_{\text{init}}$ denotes the number of initial observations collected before the sequential optimization process begins, which determines the quality of the initial surrogate model. Specifically, we consider $N_{\text{init}} = 32$ and $N_{\text{init}} = 64$ instead of $N_{\text{init}} = 12$ considered in main text.

As shown in Figure~\ref{fig:ninit_32_convergence}, when $N_{\text{init}} = 32$, all optimizers benefit from the increased number of initial observations, leading to improved overall performance. However, KSOS-BO continues to outperform the other baselines on 5 out of 6 benchmarks, indicating its robustness in relatively low-data regimes.

\begin{figure}
  \centering
  \includegraphics[width=\linewidth]{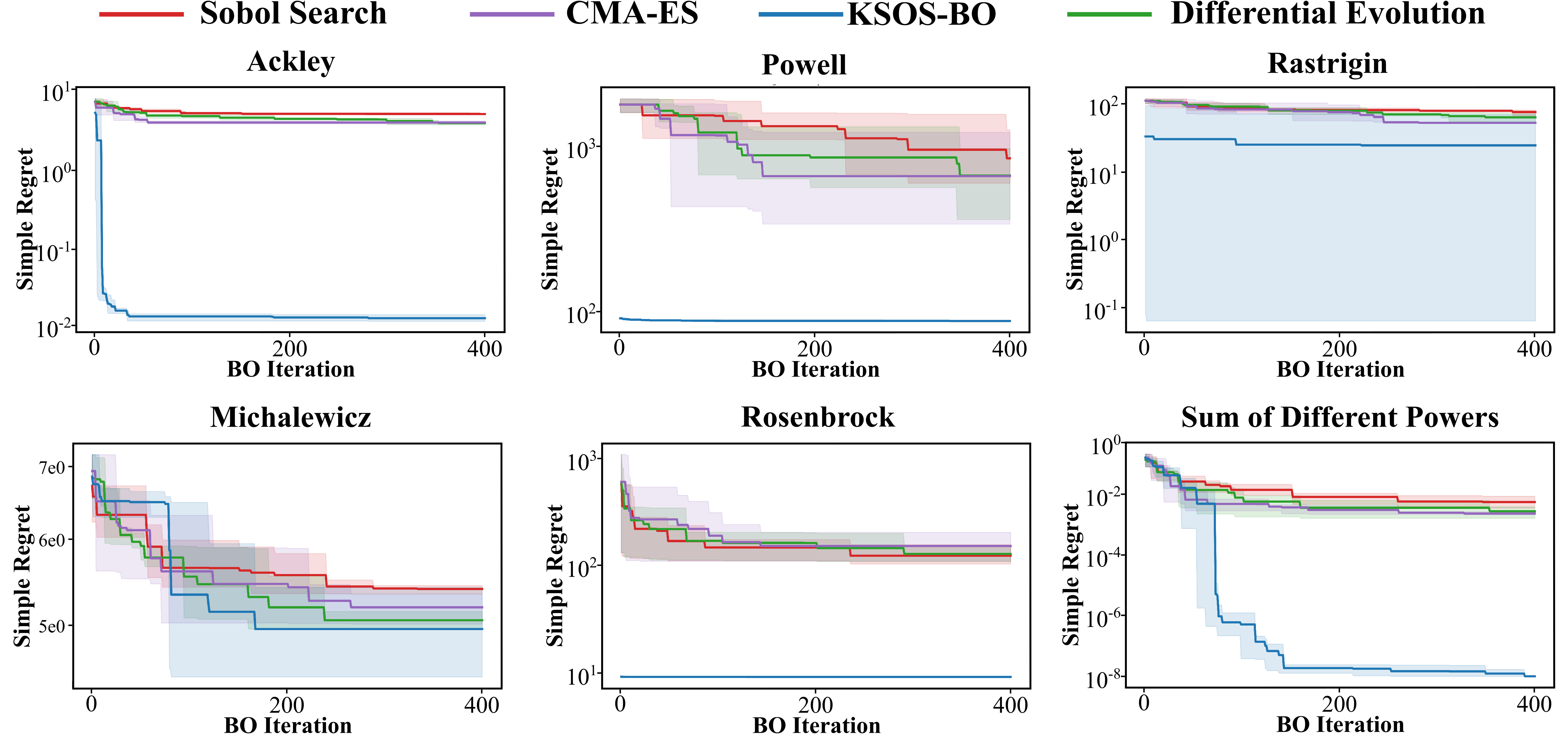}
  \caption{
Convergence comparison on 6 10-dimensional benchmark functions with $N_{\text{init}}=64$ using \ac{EI} over 400 iterations. The curves show the simple regret across 5 random seeds on a logarithmic scale, with shaded regions indicating the 95\% confidence interval.
}
\label{fig:ninit_64_convergence}
\end{figure}

As shown in Figure~\ref{fig:ninit_64_convergence}, when $N_{\text{init}} = 64$, the performance of KSOS-BO improves further, with its average rank increasing from 4 to 1 in Michalewicz. This suggests that, given a more informative initial dataset, KSOS-BO is able to more accurately approximate the acquisition landscape and identify higher-quality solutions.

\subsection{Ablation Study on Evaluation Budget Per Iteration}

$N_{\text{sample}}$ denotes the per-iteration evaluation budget. We study its impact on the performance of different optimizers.

\begin{figure}
  \centering
  \includegraphics[width=\linewidth]{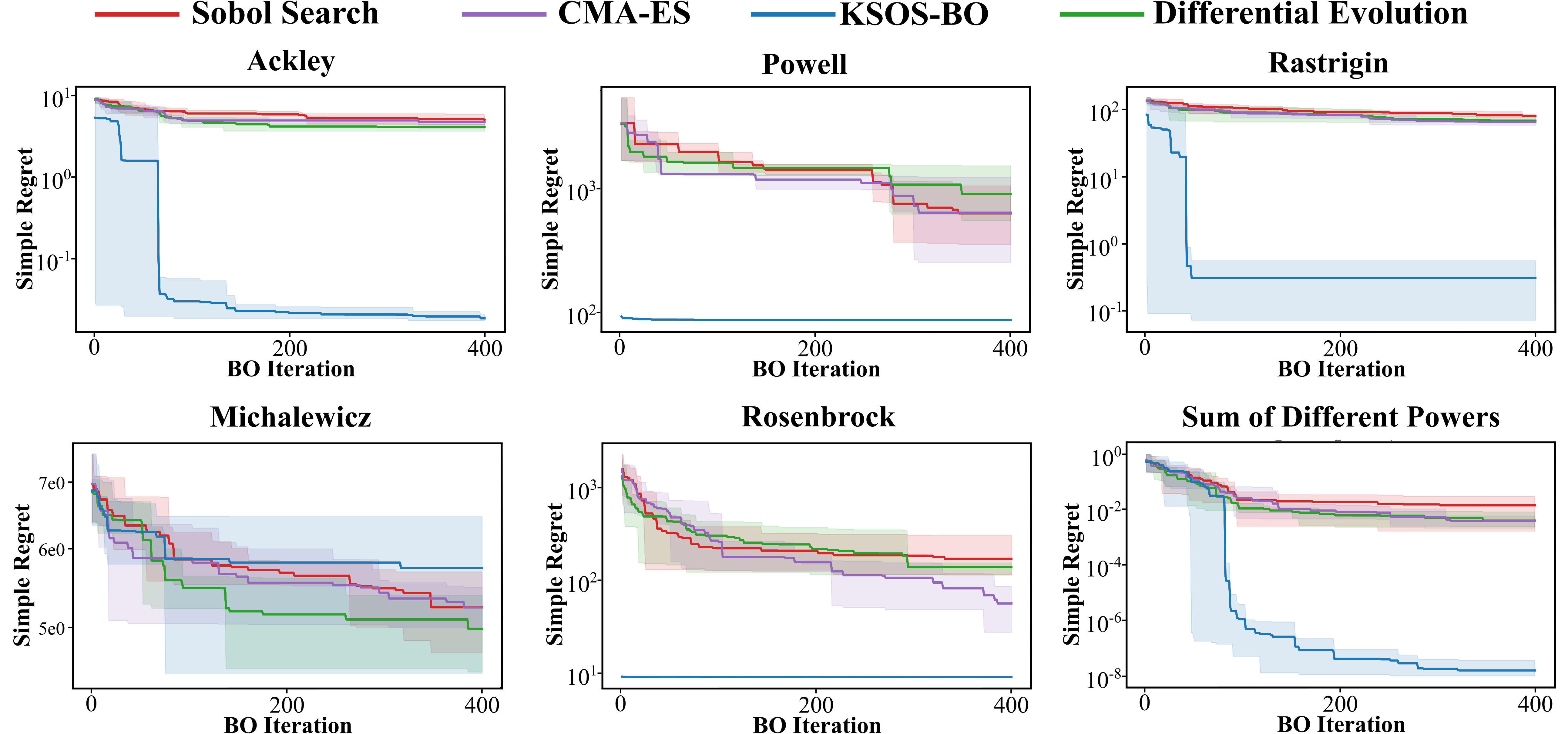}
  \caption{
Convergence comparison on 6 10-dimensional benchmark functions with $N_{\text{sample}}=64$ using \ac{EI} over 400 iterations. The curves show the simple regret across 5 random seeds on a logarithmic scale, with shaded regions indicating the 95\% confidence interval.
}
\label{fig:sample_64_convergence}
\end{figure}

As shown in Figure~\ref{fig:sample_64_convergence}, reducing $N_{\text{sample}}$ from $128$ (Figure~\ref{fig:convergence_main}) to $64$ degrades performance across all optimizers. Nevertheless, KSOS-BO still outperforms the baselines on 5 out of 6 benchmarks, indicating robustness under limited evaluation budgets.

\begin{figure}
  \centering
  \includegraphics[width=\linewidth]{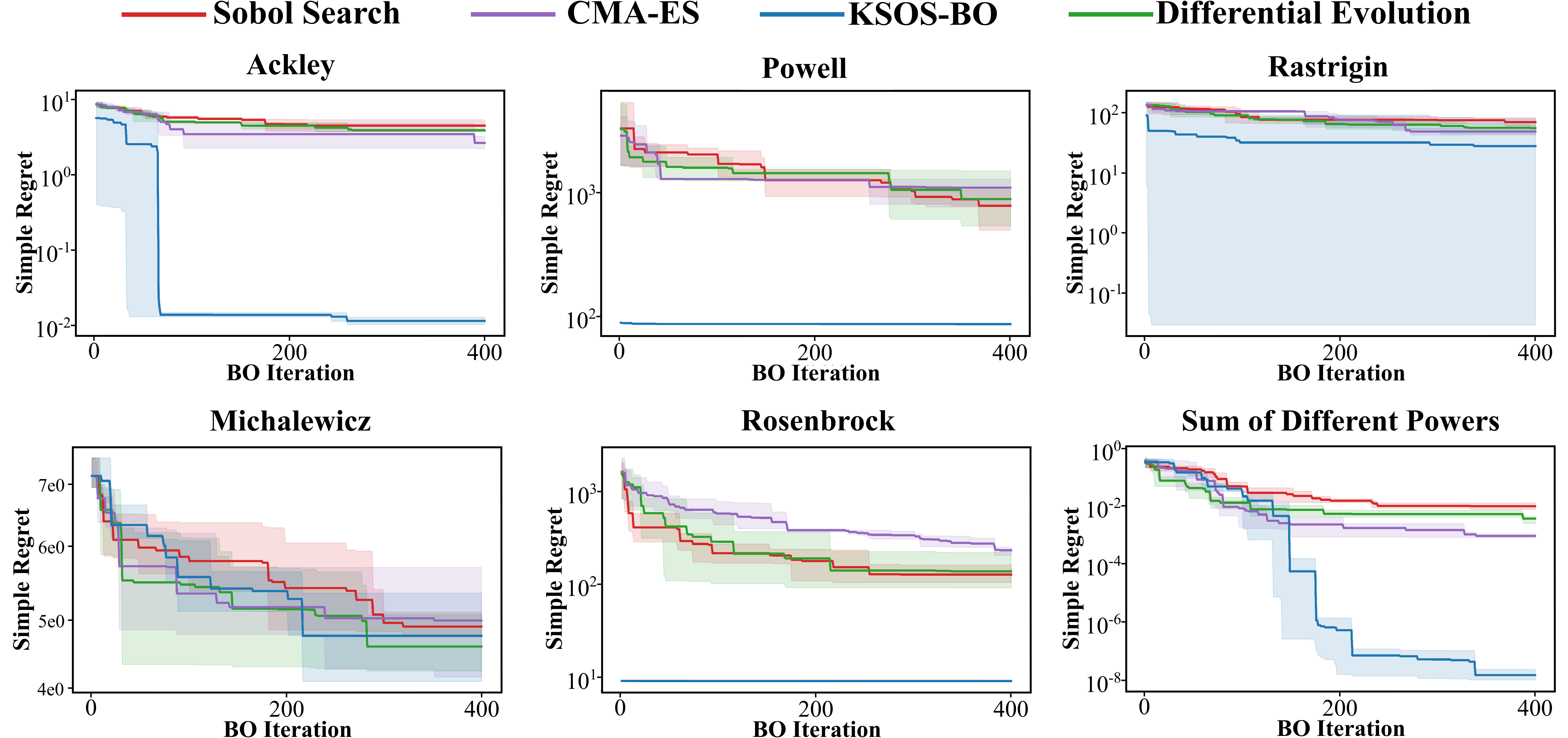}
  \caption{
Convergence comparison on 6 10-dimensional benchmark functions with $N_{\text{sample}}=256$ using \ac{EI} over 400 iterations. The curves show the simple regret across 5 random seeds on a logarithmic scale, with shaded regions indicating the 95\% confidence interval.
}
\label{fig:sample_256_convergence}
\end{figure}

As shown in Figure~\ref{fig:sample_256_convergence}, increasing $N_{\text{sample}}$ from $128$ to $256$ improves the performance of all methods. KSOS-BO maintains its advantage on 5 out of 6 benchmarks, with its performance on Michalewicz improving from rank 4 to rank 2.

Overall, KSOS-BO is less sensitive to the choice of $N_{\text{sample}}$ and effectively utilizes both limited and abundant evaluation budgets, achieving strong performance across different sampling regimes.

\subsection{Ablation Study on Dimensionality}\label{app:dim_analysis_ei}

\begin{figure}
  \centering
  \includegraphics[width=\linewidth]{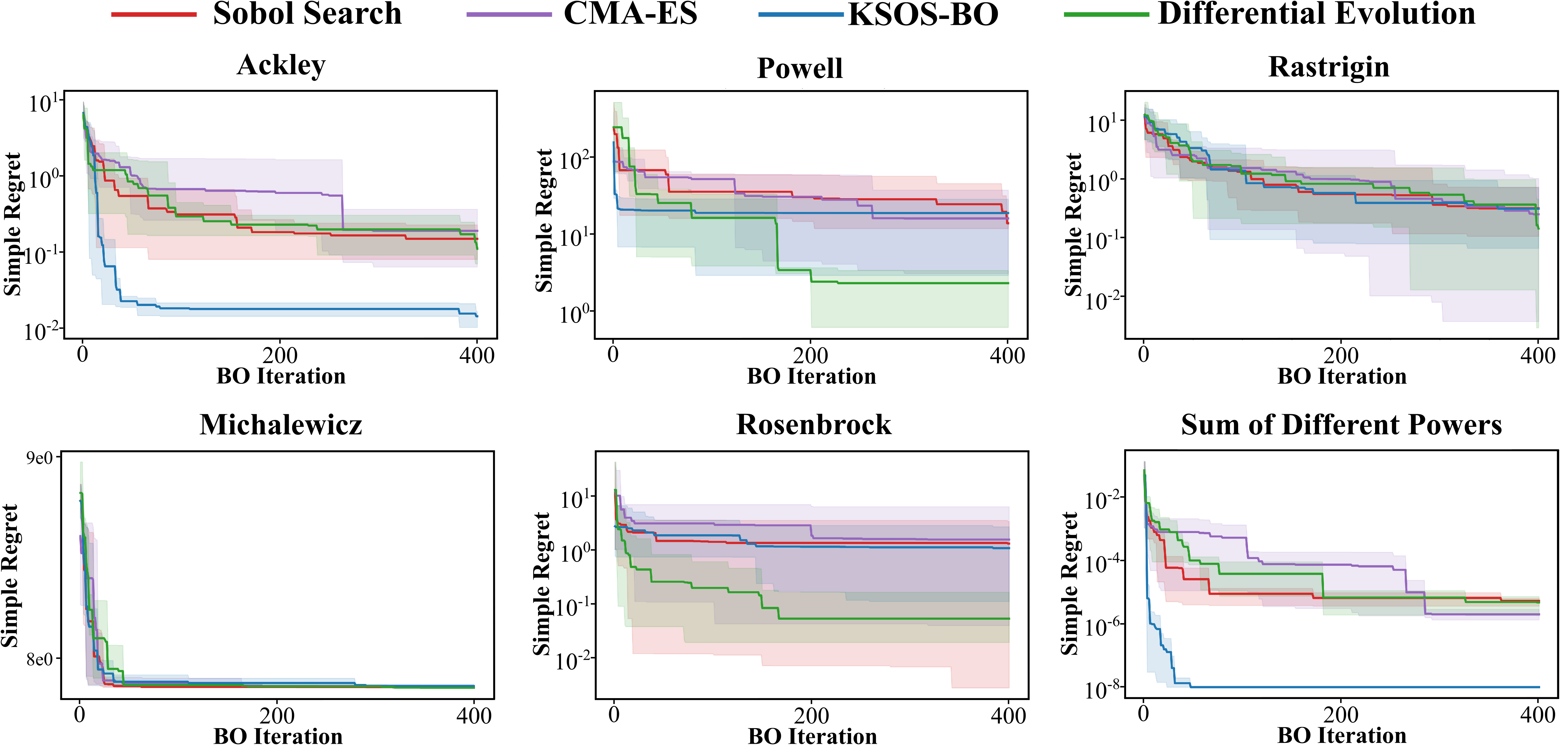}
  \caption{
Convergence comparison on 6 benchmarks in 2 dimensions using \ac{EI} over 400 iterations. The curves show the simple regret over 5 random seeds on a logarithmic scale, with shaded regions indicating the 95\% confidence interval.
}
\label{fig:dim2_ei}
\end{figure}

\begin{figure}
  \centering
  \includegraphics[width=\linewidth]{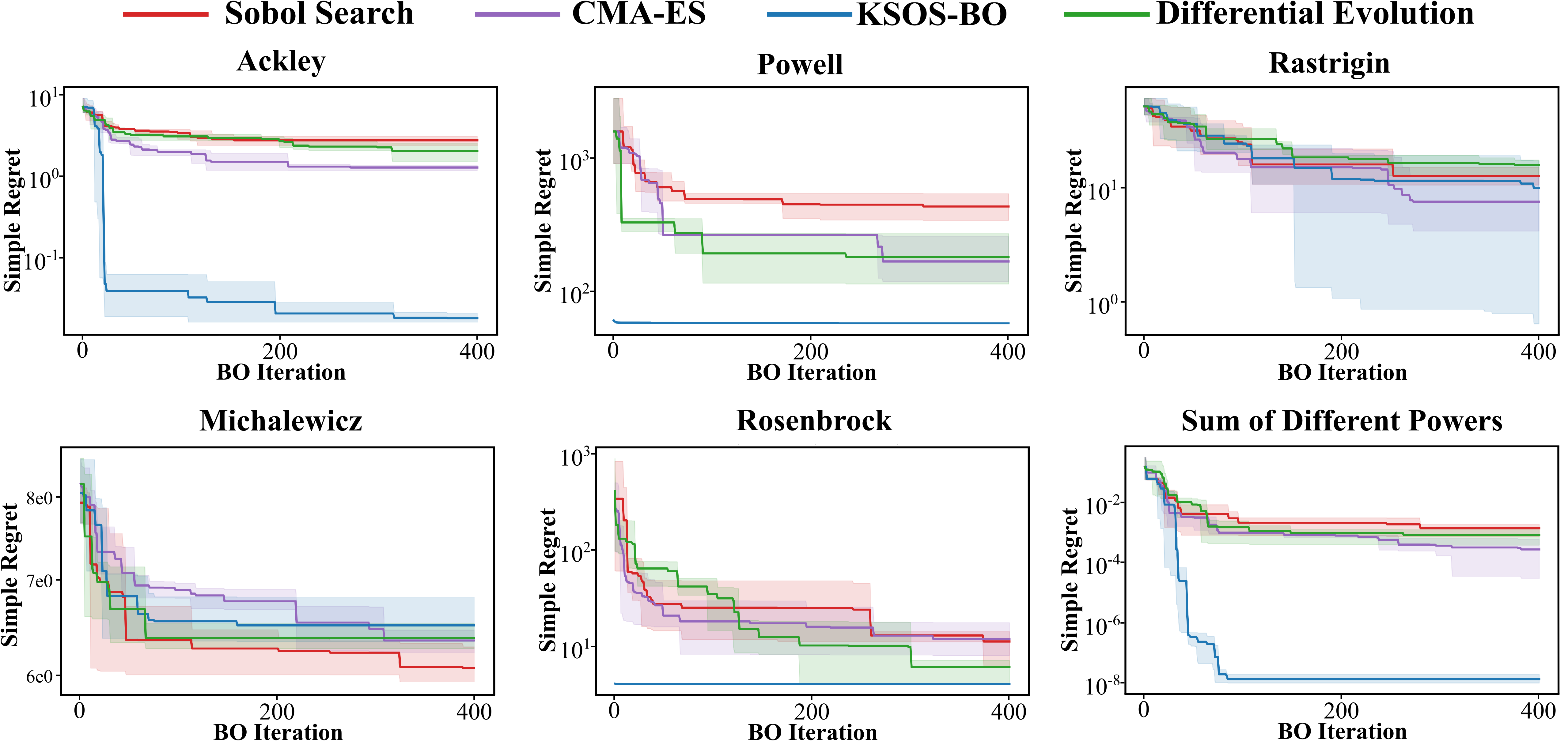}
  \caption{
Convergence comparison on 6 benchmarks in 5 dimensions using \ac{EI} over 400 iterations. The curves show the simple regret over 5 random seeds on a logarithmic scale, with shaded regions indicating the 95\% confidence interval.
}
\label{fig:dim5_ei}
\end{figure}

We further analyze the performance of all optimizers under different problem dimensions using \ac{EI} as the acquisition function. Figure~\ref{fig:dim2_ei} and Figure~\ref{fig:dim5_ei} show the mean simple regret across 5 seeds for 6 benchmarks in 2 dimensions and 5 dimensions.

In low-dimensional settings ($d=2$, and $d=4$ for Powell), KSOS-BO achieves the fastest convergence in Ackley and Sum of Different Powers, rapidly reducing regret within the first few iterations and stabilizing at significantly lower values than all baselines. In Powell, Rastrigin, and Rosenbrock, KSOS-BO performs worse than \ac{DE}, while in Michalewicz, all methods converge to similar performance. This suggests that evolutionary methods may already be sufficient in very low-dimensional settings for these functions.

In higher-dimensional settings ($d=5$, and $d=8$ for Powell), KSOS-BO continues to perform strongly on Ackley and Sum of Different Powers. Notably, on Powell and Rosenbrock, KSOS-BO improves from lower-ranked performance in $d=2$ to the top rank in $d=5$, indicating its effectiveness in higher-dimensional settings. 

Overall, these results demonstrate that KSOS-BO adapts well across different dimensionalities and exhibits strong performance even as the dimension increases, as opposed to \ac{CMA-ES}, \ac{DE} and Sobol Search, which exhibit significant performance drops.

\subsection{Ablation Study on Sampling Methods}

Here, we study the influence of sampling strategies on optimizers. We compare two simple baselines: Sobol Search and Uniform Search, both of which select the best candidate from a set of sampled points. In addition, we evaluate the effect of replacing Sobol sampling with uniform sampling within KSOS-BO. We exclude \ac{CMA-ES} and \ac{DE} from this analysis, as their sampling mechanisms are intrinsic to the algorithms and not directly comparable.

\begin{figure}
  \centering
  \includegraphics[width=\linewidth]{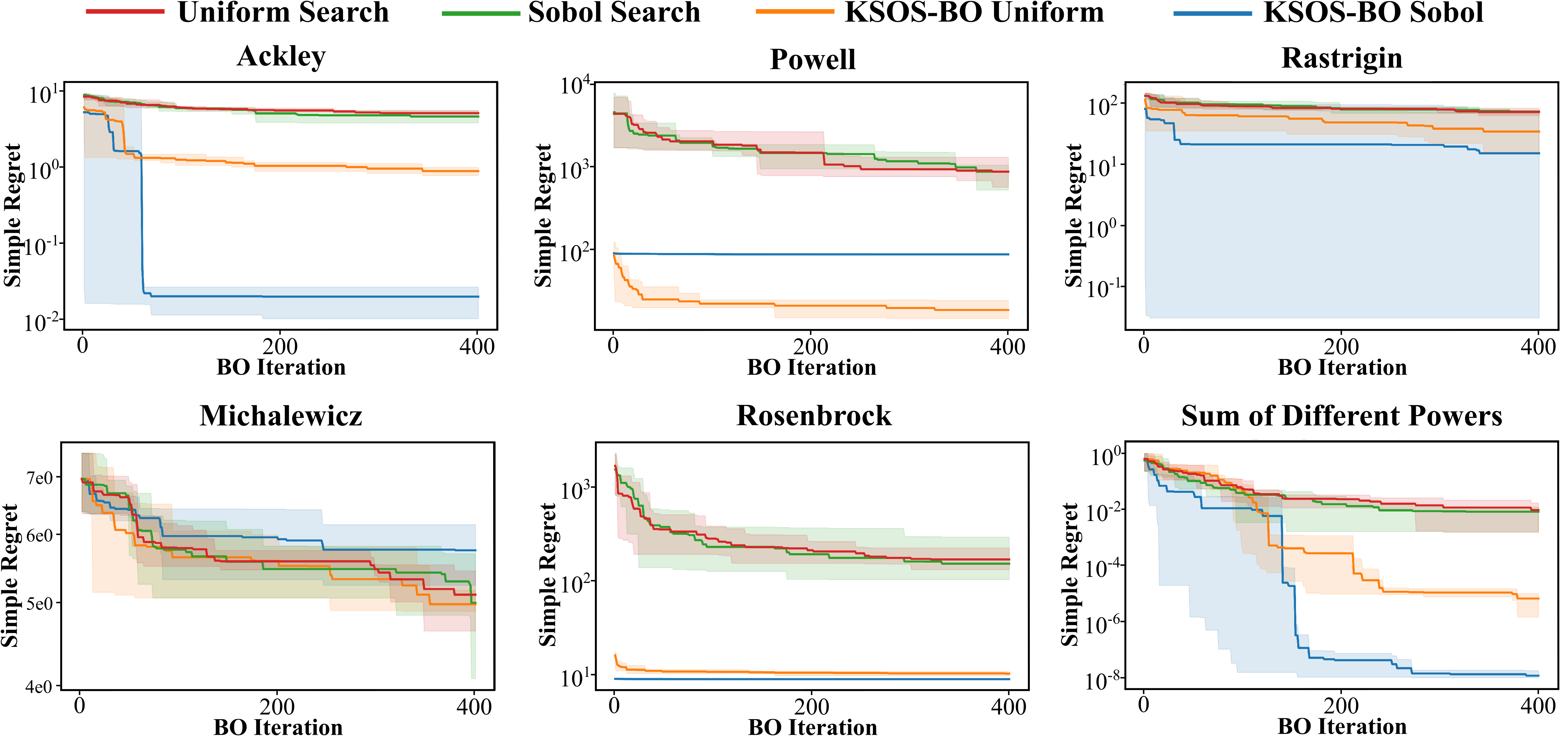}
  \caption{
Convergence comparison of Sobol and uniform sampling for KSOS-BO, along with Sobol Search and Uniform Search baselines, on 6 benchmark functions in 10 dimensions using \ac{EI} over 400 iterations. Curves show the simple regret over 5 seeds on a logarithmic scale, with shaded regions indicating the 95\% confidence interval.
}
\label{fig:sobol_uniform}
\end{figure}

As shown in Figure~\ref{fig:sobol_uniform}, Sobol Search and Uniform Search exhibit similar performance across benchmarks, indicating that sampling alone provides limited improvement when no surrogate structure is exploited. In contrast, KSOS-BO shows a pronounced sensitivity to the sampling strategy. Using Sobol sampling leads to improved performance in Ackley, Rastrigin, Rosenbrock, and Sum of Different Powers, while uniform sampling performs better in Michalewicz and Powell.

This behavior can be explained by the fact that \ac{KernelSOS} constructs its surrogate entirely from the evaluations of the samples of functions. Consequently, the distribution of sampled points directly affects the quality of the fitted surrogate and optimization performance.

\subsection{Ablation Study on Population Size in \ac{CMA-ES}}\label{app:cmaes}

We study the influence of population size on the performance of \ac{CMA-ES}. Specifically, we evaluate 3 population sizes, $8$, $16$, and $32$, and compare them with KSOS-BO under the same evaluation budget of $128$.

\begin{figure}
  \centering
  \includegraphics[width=\linewidth]{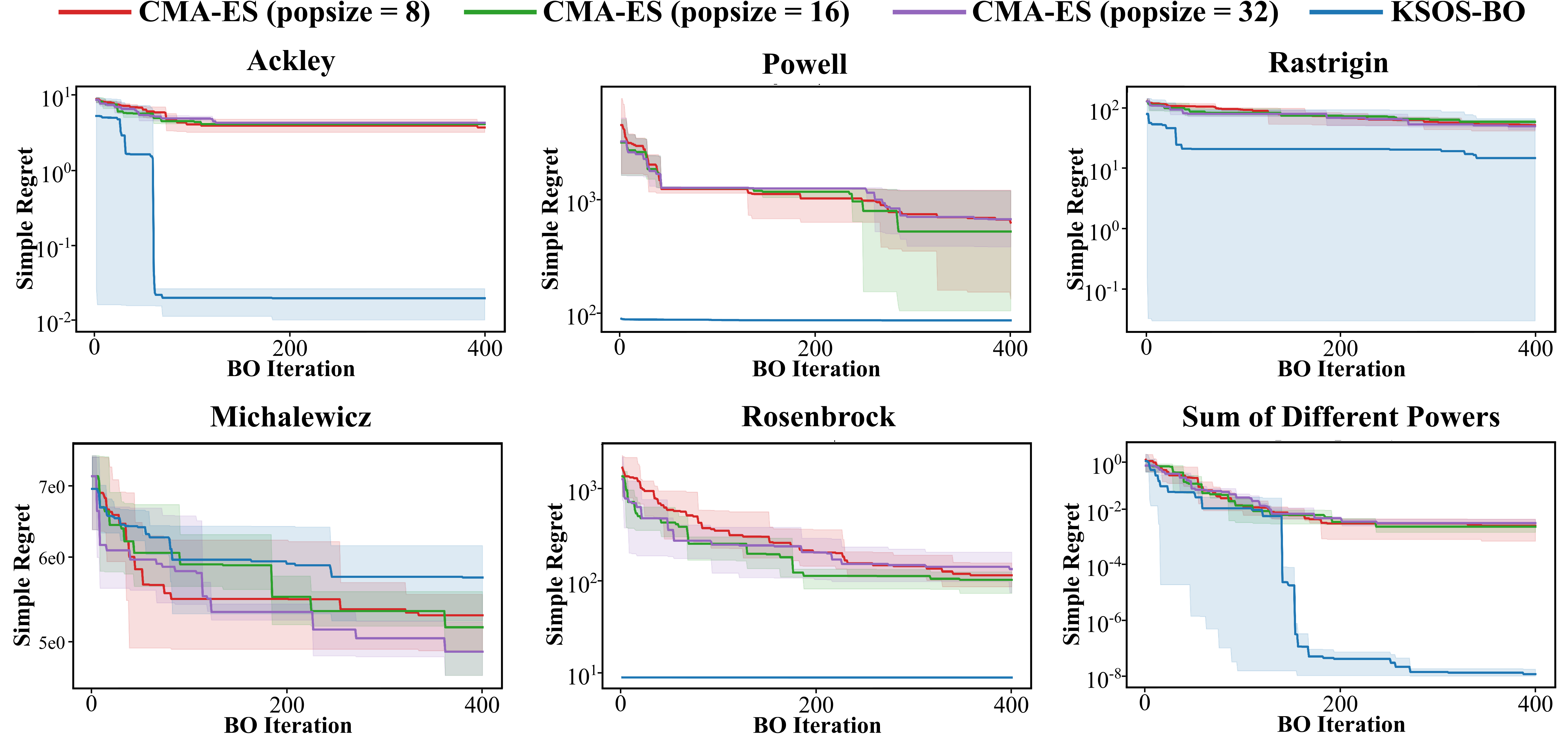}
  \caption{
Convergence comparison of \ac{CMA-ES} with different population sizes to KSOS-BO on 6 10-dimensional benchmark functions using \ac{EI} over 400 iterations. Curves show the simple regret across 5 seeds on a logarithmic scale, with shaded regions indicating the 95\% confidence interval.
}
\label{fig:cmaes}
\end{figure}

As shown in Figure~\ref{fig:cmaes}, increasing the population size from 8 to 16 improves the performance of \ac{CMA-ES} on Powell and Rosenbrock. Further increasing the population size from 16 to 32 leads to improvements on the Michalewicz function. On Ackley, Rastrigin, and Sum of Different Powers, \ac{CMA-ES} exhibits similar performance across different population sizes, with only minor variations. In all cases, KSOS-BO outperforms \ac{CMA-ES} on 5 out of 6 benchmarks, which is consistent with the result shown in Figure~\ref{fig:convergence_main}. This suggests that the observed performance gap of \ac{CMA-ES} and KSOS-BO is insensitive to the choice of population size.

\subsection{Ablation Study on Population Size and Population Diversity in \ac{DE}}\label{app:de}

For \ac{DE}, we consider two additional configurations under the same evaluation budget compared to the setting used in the main text. The first increases the population size from $2$ to $3$ while reducing the number of generations from $5$ to $3$, testing whether greater population diversity is more beneficial than additional iterative refinement. The second increases the crossover rate from $0.7$ to $0.9$, testing whether stronger coordinate mixing improves optimization performance.

\begin{figure}
  \centering
  \includegraphics[width=\linewidth]{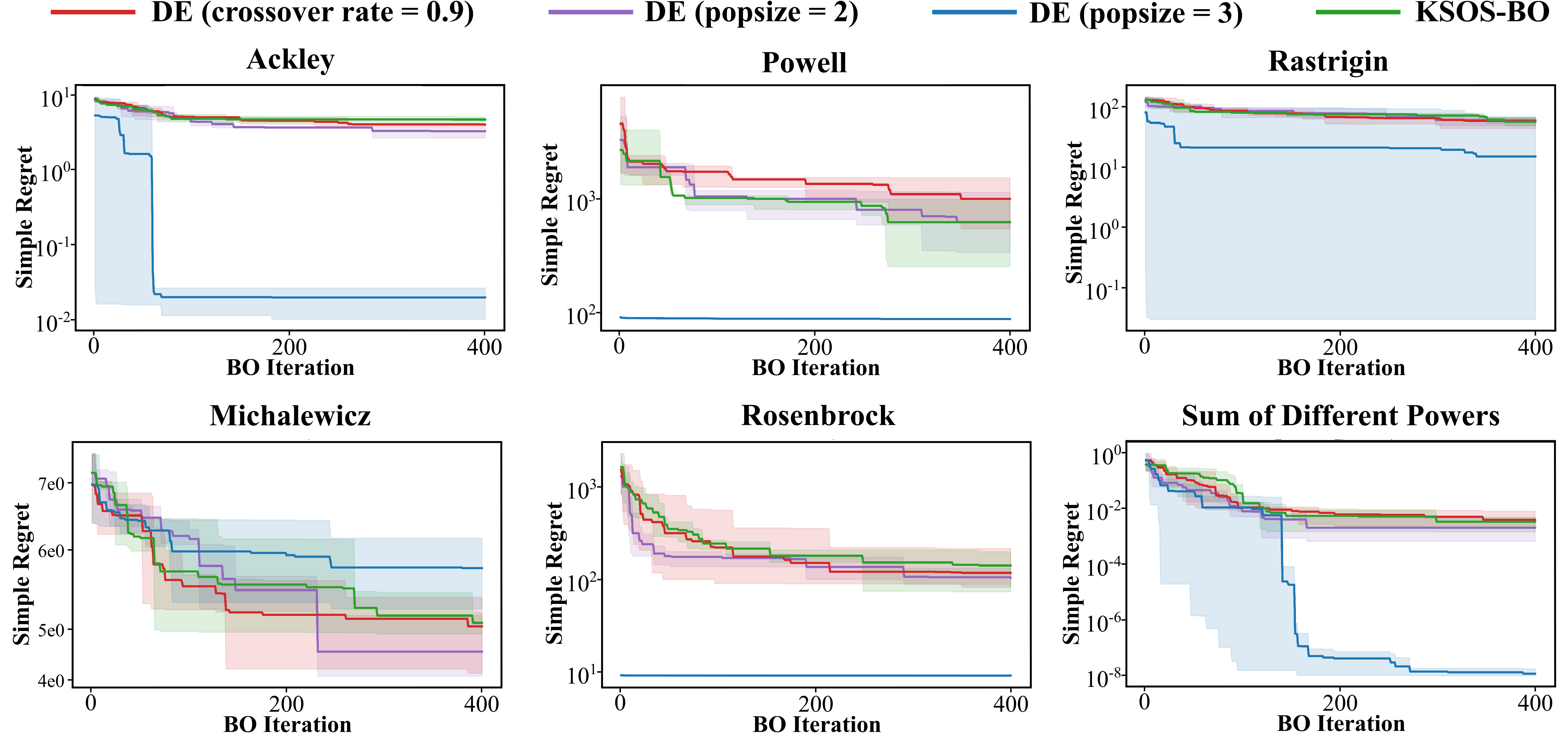}
  \caption{
Convergence comparison of \ac{DE} with different population sizes and crossover rate to KSOS-BO on 6 10-dimensional benchmark functions using \ac{EI} over 400 iterations. Curves show the simple regret across 5 seeds on a logarithmic scale, with shaded regions indicating the 95\% confidence interval.
}
\label{fig:de_ablation}
\end{figure}

As shown in Figure~\ref{fig:de_ablation}, increasing the population size improves \ac{DE} performance on Powell compared to the setting used in the main text. Increasing the crossover rate improves performance on Ackley, Michalewicz, Rosenbrock and Sum of Different Powers. However, in all cases, KSOS-BO outperforms \ac{DE} on 5 out of 6 benchmarks, which is consistent with the result shown in Figure~\ref{fig:convergence_main}. These results suggest that the observed performance gap between \ac{DE} and KSOS-BO is insensitive to the choice of population size and crossover rate.

\subsection{Ablation Study on KSOS-BO with Laplace Kernel}\label{app:laplace}

\begin{figure}
  \centering
  \includegraphics[width=\linewidth]{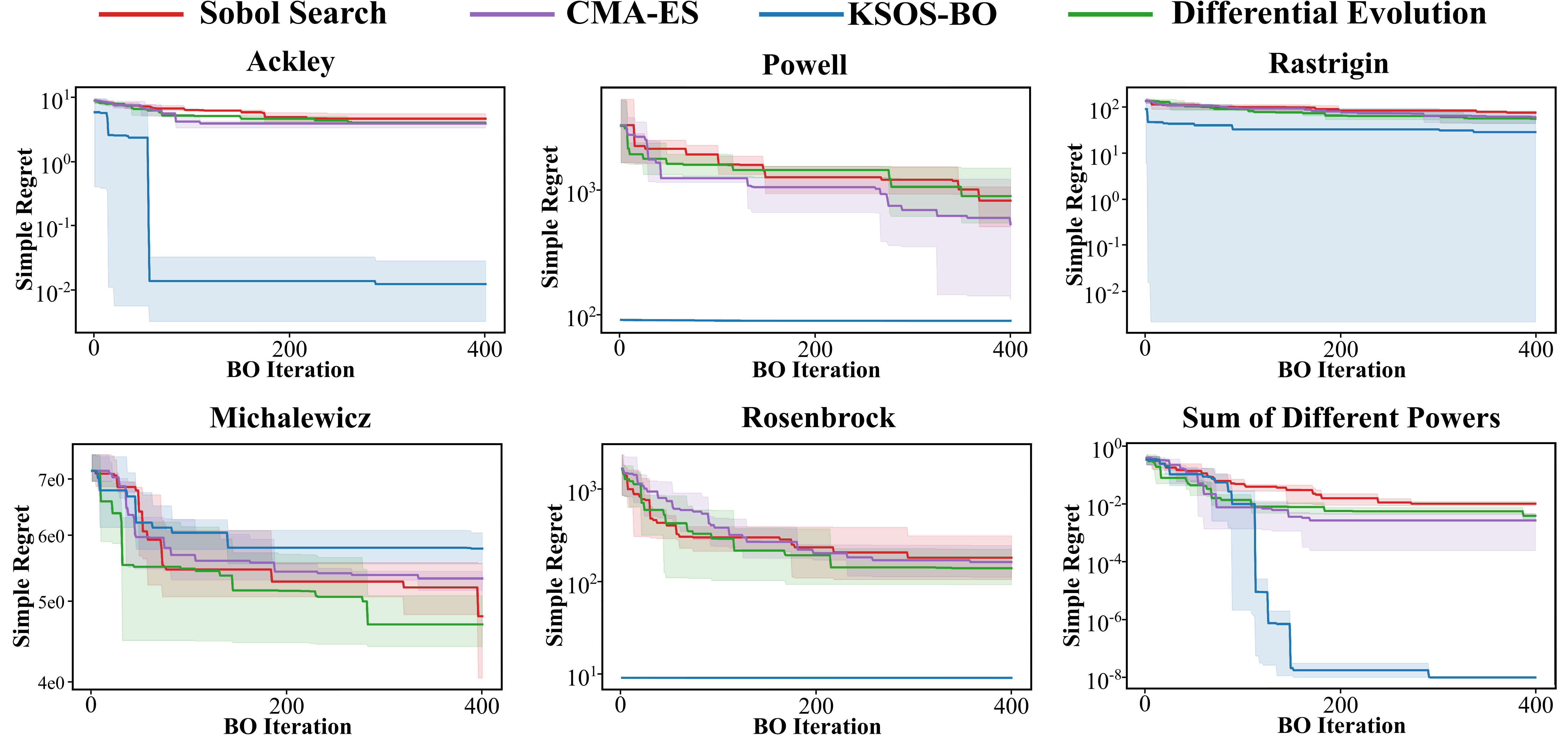}
  \caption{
Convergence comparison on 6 benchmark functions in 10 dimensions using \ac{EI} over 400 iterations, with KSOS-BO using a Laplace kernel. Curves show the simple regret across 5 seeds on a logarithmic scale, with shaded regions indicating the 95\% confidence interval.
}
\label{fig:laplace_ksos}
\end{figure}

We study the effect of using a Laplace kernel in KSOS-BO. As shown in Figure~\ref{fig:laplace_ksos}, the results are consistent with those obtained using a Gaussian kernel~\ref{fig:convergence_main}. With the Laplace kernel, KSOS-BO shows slight improvements on Ackley and Rastrigin, and exhibits lower variance on Sum of Different Powers. Overall, KSOS-BO continues to outperform the baselines on 5 out of 6 benchmarks, indicating that its performance is robust to the choice of kernel.

\subsection{Wall-clock Time Improvement}\label{app:runtime_table}

To summarize the runtime efficiency of KSOS-BO across benchmark functions, we report both the final ranking and the relative time-to-threshold improvement in Table~\ref{tab:ksos_summary_time}. For each benchmark, we first compute the final aggregated regret of each acquisition optimizer at the last optimization step and rank the optimizers in ascending order, since lower regret indicates better performance.

When KSOS-BO achieves the best final performance, we use the second-best optimizer as the reference. Let $r_{\mathrm{ref}}$ and $T_{\mathrm{ref}}$ denote the final regret and final runtime of the reference optimizer, respectively. We then define the earliest runtime at which KSOS-BO reaches the same regret level as

\begin{equation}
T_K
=
\inf \left\{
t : r_K(t) \leq r_{\mathrm{ref}}
\right\},
\end{equation}

where $T_K$ is the first hitting time at which the regret of KSOS-BO falls below the reference level $r_{\mathrm{ref}}$, $r_K(t)$ denotes the simple regret of KSOS-BO at runtime $t$.
The relative runtime improvement is computed as
\begin{equation}
\mathrm{Improvement}(\%)
=
\frac{
T_{\mathrm{ref}} - T_K
}{
T_{\mathrm{ref}}
}
\times 100.
\end{equation}
This metric measures how much earlier KSOS-BO reaches the target regret defined by the strongest competing optimizer. A positive value indicates faster convergence in wall-clock time.

Table~\ref{tab:ksos_summary_time} shows that KSOS-BO ranks first on 10 out of 15 benchmark functions, with particularly strong performance on multimodal, bowl-shaped, and valley-shaped landscapes. The improvement is especially pronounced on Ackley, Levy, Griewank, Rotated Hyper Ellipsoid, Trid, Dixon Price, Rosenbrock, and Powell. These results indicate that KSOS-BO improves not only the final optimization quality but also wall-clock efficiency.

In several functions, including Schwefel, Michalewicz, Sphere, Zakharov, and Styblinski Tang, KSOS-BO does not achieve the top rank. These cases suggest that the effectiveness of KSOS-BO depends on the structure of the acquisition landscape and that extremely flat, steep, or irregular landscapes may diminish the advantage of the proposed method.

\begin{table}[t]
\caption{Summary of KSOS-BO performance across benchmark functions in wall-clock time. The table reports the relative ranking and percentage improvement compared to the second-best optimizer.}
\label{tab:ksos_summary_time}
\centering
\small 
\begin{tabular}{llcccc}
\toprule
\textbf{Type} & \textbf{Function} & \textbf{KSOS-BO Rank} & \textbf{Improvement (\%)} \\
\midrule

\multirow{5}{*}{Multimodal}
 & Ackley        & \textbf{1} & 89.15\\
 & Rastrigin     & \textbf{1} & 87.47\\
 & Levy        & \textbf{1}    & 99.31\\
 & Griewank        & \textbf{1}    & 99.59\\
 & Schwefel        & 4    & -31.88\\

\midrule

\multirow{1}{*}{Steep Drops}
 & Michalewicz    & 4   & -6.74\\

\midrule

\multirow{4}{*}{Bowl Shaped}
 & Rotated Hyper Ellipsoid        & \textbf{1}   & 99.64\\
 & Sphere     & 2    & -115.99\\
 & Sum of Different Powers      & \textbf{1}   & 61.75\\
 & Trid      & \textbf{1}  & 99.66\\

\midrule

\multirow{1}{*}{Plate Shaped}
 & Zakharov    & 4  & -1944.61\\

\midrule

\multirow{2}{*}{Valley Shaped}
 & Dixon Price    & \textbf{1}   & 99.57\\
 & Rosenbrock    & \textbf{1}  & 99.67\\

\midrule

\multirow{2}{*}{Other}
 & Powell    & \textbf{1}   & 99.64\\
 & Styblinski Tang    & 4  & -115.21\\

\bottomrule
\end{tabular}
\end{table}

\subsection{Hardware and Operating System}

All experiments are conducted on a machine equipped with an AMD Ryzen 7 5800H CPU (16 threads) and an NVIDIA GeForce RTX 3060 GPU. The system runs Ubuntu 20.04.6 LTS with Linux kernel version 5.15.0-139-generic on an x86\_64 architecture. All experiments are implemented in Python 3.12.12, ensuring a consistent computational environment across all benchmarks. As for the environment, we use NumPy, SciPy, scikit-learn, and CMA-ES (via the cma library~\cite{nomura2026cmaes}) as the underlying optimization and modeling frameworks for all benchmark algorithms. \ac{KernelSOS} is implemented using~\cite{ksos_tools}.

\subsection{LLM Usage Statement}
We acknowledge the use of LLMs (ChatGPT, Claude, and Gemini) only for polishing the writing of this paper.

\end{document}

%% file: acronyms.tex
\acrodef{TFM}[TFM]{Tabular Foundation Models}
\acrodef{CMA-ES}[CMA-ES]{Covariance Matrix Adaptation Evolution Strategy}
\acrodef{DE}[DE]{Differential Evolution}
\acrodef{BO}[BO]{Bayesian Optimization}
\acrodef{SOS}[SOS]{Sum-of-Squares}
\acrodef{KernelSOS}[KernelSOS]{Kernel Sum-of-Squares}
\acrodef{RKHS}[RKHS]{Reproducing Kernel Hilbert Spaces}
\acrodef{EI}[EI]{Expected Improvement}
\acrodef{LCB}[LCB]{Lower Bound Confidence}
\acrodef{LMI}[LMI]{Linear Matrix Inequality}
\acrodef{GP}[GP]{Gaussian Process}

%% file: main.bbl
\begin{thebibliography}{10}

\bibitem{shahriari2015taking}
Bobak Shahriari, Kevin Swersky, Ziyu Wang, Ryan~P Adams, and Nando De~Freitas.
\newblock {Taking the human out of the loop: A review of Bayesian optimization}.
\newblock {\em Proceedings of the IEEE}, 104(1):148--175, 2015.

\bibitem{snoek2015scalable}
Jasper Snoek, Oren Rippel, Kevin Swersky, Ryan Kiros, Nadathur Satish, Narayanan Sundaram, Mostofa Patwary, Prabhat, and Ryan~P. Adams.
\newblock {Scalable Bayesian Optimization Using Deep Neural Networks}.
\newblock In {\em Proceedings of the International Conference on Machine Learning}, pages 2171--2180. PMLR, 2015.

\bibitem{wang2017new}
Hao Wang, Bas van Stein, Michael Emmerich, and Thomas Back.
\newblock {A new acquisition function for Bayesian optimization based on the moment-generating function}.
\newblock In {\em 2017 IEEE International Conference on Systems, Man, and Cybernetics (SMC)}, pages 507--512. IEEE, 2017.

\bibitem{gan2021acquisition}
Weiao Gan, Ziyuan Ji, and Yongqing Liang.
\newblock {Acquisition functions in Bayesian optimization}.
\newblock In {\em 2021 2nd International Conference on Big Data \& Artificial Intelligence \& misc Engineering (ICBASE)}, pages 129--135. IEEE, 2021.

\bibitem{wilson2018maximizing}
James Wilson, Frank Hutter, and Marc Deisenroth.
\newblock {Maximizing acquisition functions for Bayesian optimization}.
\newblock {\em Advances in Neural Information Processing Systems}, 31, 2018.

\bibitem{das2010differential}
Swagatam Das and Ponnuthurai~Nagaratnam Suganthan.
\newblock {Differential Evolution: A survey of the state-of-the-art}.
\newblock {\em IEEE Transactions on Evolutionary Computation}, 15(1):4--31, 2010.

\bibitem{hansen2016cma}
Nikolaus Hansen.
\newblock {The CMA evolution strategy: A tutorial}.
\newblock {\em arXiv preprint arXiv:1604.00772}, 2016.

\bibitem{eggensperger2015efficient}
Katharina Eggensperger, Frank Hutter, Holger Hoos, and Kevin Leyton-Brown.
\newblock {Efficient benchmarking of hyperparameter optimizers via surrogates}.
\newblock In {\em Proceedings of the AAAI Conference on Artificial Intelligence}, volume~29, 2015.

\bibitem{nocedal2006numerical}
Jorge Nocedal and Stephen~J Wright.
\newblock {\em {Numerical optimization}}.
\newblock Springer, 2006.

\bibitem{lasserre2001global}
Jean~B Lasserre.
\newblock {Global optimization with polynomials and the problem of moments}.
\newblock {\em SIAM Journal on Optimization}, 11(3):796--817, 2001.

\bibitem{parrilo2003semidefinite}
Pablo~A Parrilo.
\newblock {Semidefinite programming relaxations for semialgebraic problems}.
\newblock {\em Mathematical Programming}, 96(2):293--320, 2003.

\bibitem{lasserre2009moments}
Jean~Bernard Lasserre.
\newblock {\em {Moments, positive polynomials and their applications}}, volume~1.
\newblock World Scientific, 2009.

\bibitem{snoek2012practical}
Jasper Snoek, Hugo Larochelle, and Ryan~P Adams.
\newblock {Practical bayesian optimization of machine learning algorithms}.
\newblock {\em Advances in Neural Information Processing Systems}, 25, 2012.

\bibitem{rudi2025finding}
Alessandro Rudi, Ulysse Marteau-Ferey, and Francis Bach.
\newblock {Finding global minima via kernel approximations: A. Rudi et al.}
\newblock {\em Mathematical Programming}, 209(1):703--784, 2025.

\bibitem{parrilo2000structured}
Pablo~A Parrilo.
\newblock {\em {Structured Semidefinite Programs and Semialgebraic Geometry Methods in Robustness and Optimization}}.
\newblock PhD thesis, California Institute of Technology, 2000.

\bibitem{tedrake2010lqr}
Russ Tedrake, Ian~R Manchester, Mark Tobenkin, and John~W Roberts.
\newblock {LQR-trees: Feedback motion planning via sums-of-squares verification}.
\newblock {\em The International Journal of Robotics Research}, 29(8):1038--1052, 2010.

\bibitem{yang2022certifiably}
Heng Yang and Luca Carlone.
\newblock {Certifiably optimal outlier-robust geometric perception: Semidefinite relaxations and scalable global optimization}.
\newblock {\em IEEE Transactions on Pattern Analysis and Machine Intelligence}, 45(3):2816--2834, 2022.

\bibitem{lasserre2015introduction}
Jean~Bernard Lasserre.
\newblock {\em {An introduction to polynomial and semi-algebraic optimization}}, volume~52.
\newblock Cambridge University Press, 2015.

\bibitem{paulsen2016introduction}
Vern~I Paulsen and Mrinal Raghupathi.
\newblock {\em {An introduction to the theory of reproducing kernel Hilbert spaces}}, volume 152.
\newblock Cambridge University Press, 2016.

\bibitem{scholkopf2018learning}
Bernhard Sch{\"o}lkopf and Alexander~J. Smola.
\newblock {\em {Learning with Kernels: Support Vector Machines, Regularization, Optimization, and Beyond}}.
\newblock MIT Press, 2018.

\bibitem{berthier2022infinite}
Elo{\"\i}se Berthier, Justin Carpentier, Alessandro Rudi, and Francis Bach.
\newblock Infinite-dimensional sums-of-squares for optimal control.
\newblock In {\em 2022 IEEE 61st Conference on Decision and Control (CDC)}, pages 577--582. IEEE, 2022.

\bibitem{lasserre2008nonlinear}
Jean~B Lasserre, Didier Henrion, Christophe Prieur, and Emmanuel Tr{\'e}lat.
\newblock {Nonlinear optimal control via occupation measures and LMI-relaxations}.
\newblock {\em SIAM Journal on Control and Optimization}, 47(4):1643--1666, 2008.

\bibitem{groudiev2025sampling}
Antoine Groudiev, Fabian Schramm, {\'E}lo{\"\i}se Berthier, Justin Carpentier, and Frederike D{\"u}mbgen.
\newblock Sampling-based global optimal control and estimation via semidefinite programming.
\newblock {\em arXiv preprint arXiv:2507.17572}, 2025.

\bibitem{wei2026global}
Zhongqi Wei and Frederike D{\"u}mbgen.
\newblock Global sampling-based trajectory optimization for contact-rich manipulation via kernelsos.
\newblock {\em arXiv preprint arXiv:2604.27175}, 2026.

\bibitem{frazier2018bayesian}
Peter~I Frazier.
\newblock {Bayesian optimization}.
\newblock In {\em Recent Advances in Optimization and Modeling of Contemporary Problems}, pages 255--278. Informs, 2018.

\bibitem{jones1998efficient}
Donald~R Jones, Matthias Schonlau, and William~J Welch.
\newblock {Efficient global optimization of expensive black-box functions}.
\newblock {\em Journal of Global optimization}, 13(4):455--492, 1998.

\bibitem{bergstra2011algorithms}
James Bergstra, R{\'e}mi Bardenet, Yoshua Bengio, and Bal{\'a}zs K{\'e}gl.
\newblock {Algorithms for hyper-parameter optimization}.
\newblock {\em Advances in Neural Information Processing Systems}, 24, 2011.

\bibitem{conn2009introduction}
Andrew~R Conn, Katya Scheinberg, and Luis~N Vicente.
\newblock {\em {Introduction to derivative-free optimization}}.
\newblock SIAM, 2009.

\bibitem{brochu2009tutorial}
Eric Brochu, Vlad~M. Cora, and Nando de~Freitas.
\newblock {A Tutorial on Bayesian Optimization of Expensive Cost Functions, with Application to Active User Modeling and Hierarchical Reinforcement Learning}.
\newblock Technical report, Department of Computer Science, University of British Columbia, 2009.

\bibitem{seeger2004gaussian}
Matthias Seeger.
\newblock {Gaussian processes for machine learning}.
\newblock {\em International Journal of Neural Systems}, 14(02):69--106, 2004.

\bibitem{marteau2020non}
Ulysse Marteau-Ferey, Francis Bach, and Alessandro Rudi.
\newblock {Non-parametric models for non-negative functions}.
\newblock {\em Advances in Neural Information Processing Systems}, 33:12816--12826, 2020.

\bibitem{simulationlib}
S.~Surjanovic and D.~Bingham.
\newblock {Virtual Library of Simulation Experiments: Test Functions and Datasets}.
\newblock \url{http://www.sfu.ca/~ssurjano}, 2026.
\newblock Accessed: May 6, 2026.

\bibitem{williams1995gaussian}
Christopher Williams and Carl Rasmussen.
\newblock {Gaussian processes for regression}.
\newblock {\em Advances in Neural Information Processing Systems}, 8, 1995.

\bibitem{ksos_tools}
Antoine Groudiev and Frederike D{\"u}mbgen.
\newblock {{ksos-tools}: Implementation of the {Kernel Sum-of-Squares} optimization framework and various related solvers}.
\newblock \url{https://github.com/Simple-Robotics/ksos-tools}, 2025.
\newblock Version 0.2.2.

\bibitem{nomura2026cmaes}
Masahiro Nomura, Masashi Shibata, and Ryoki Hamano.
\newblock {cmaes: A Simple yet Practical Python Library for CMA-ES}.
\newblock {\em arXiv preprint arXiv:2402.01373}, 2026.

\end{thebibliography}
